\documentclass[12pt]{colt}

\title[Community Detection via Adaptive Sampling]{Community Detection via Random and Adaptive Sampling}
\usepackage{times}

\coltauthor{\Name{Se-Young Yun} \Email{seyoung@kth.se}\\
\Name{Alexandre Proutiere} \Email{alepro@kth.se}\\
\addr KTH, Osquldasv. 10, plan 6, 100-44, Stockholm, Sweden}

\usepackage{amsmath,amssymb, amsfonts}
\usepackage{bm}
\usepackage{algorithm,algorithmic,bbm}

\newcommand{\Ex}{{\mathbb E}}

\begin{document} 

\maketitle

\begin{abstract}
In this paper, we consider networks consisting of a finite number of non-overlapping communities. To extract these communities, the interaction between pairs of nodes may be sampled from a large available data set, which allows a given node pair to be sampled several times. When a node pair is sampled, the observed outcome is a binary random variable, equal to 1 if nodes interact and to 0 otherwise. The outcome is more likely to be positive if nodes belong to the same communities. For a given budget of node pair samples or observations, we wish to jointly design a sampling strategy (the sequence of sampled node pairs) and a clustering algorithm that recover the hidden communities with the highest possible accuracy. We consider both non-adaptive and adaptive sampling strategies, and for both classes of strategies, we derive fundamental performance limits satisfied by any sampling and clustering algorithm. In particular, we provide necessary conditions for the existence of algorithms recovering the communities accurately as the network size grows large. We also devise simple algorithms that accurately reconstruct the communities when this is at all possible, hence proving that the proposed necessary conditions for accurate community detection are also sufficient. The classical problem of community detection in the stochastic block model can be seen as a particular instance of the problems consider here. But our framework covers more general scenarios where the sequence of sampled node pairs can be designed in an adaptive manner. The paper provides new results for the stochastic block model, and extends the analysis to the case of adaptive sampling.
\end{abstract}

\section{Introduction}\label{sec:introduction}

Extracting structures or communities in networks is a central task in many disciplines including social sciences, biology, computer science, statistics, and physics. Applications are numerous. For instance, in social networks, one hopes that identifying clusters of users provides fundamental insights into the way users behave and interact, and in turn, helps the design of efficient recommender systems or the development of other marketing and advertisement techniques. Naturally, a user is attached to a particular community if she interacts a lot more with users within this cluster than with other users. Most methods for community detection assume that user interactions can be represented as a graph whose edges represent user pairs known to interact. This graph is first extracted from observed pairwise interactions and then partitioned into communities. Hence in most studies, the process of gathering information on users and the extraction of communities are decoupled. 

In this paper, we address the problems of gathering information and clustering jointly. The interaction between pairs of nodes may be sampled from a large available data set, which allows a given node pair to be sampled several times. When a node pair is sampled, the observed outcome is a binary random variable, equal to 1 if nodes interact and to 0 otherwise. Observing an interaction is more likely when nodes belong to the same community than when they don't. For a given budget of node pair samples or observations, we wish to jointly design a sampling strategy (the sequence of sampled node pairs) and a clustering algorithm that recovers the hidden communities with the highest possible accuracy, i.e., the proportion of misclassified nodes has to be minimized. We investigate two classes of sampling strategies: (i) non-adaptive random strategies where the sequence of observed node pairs is decided a priori, and (ii) adaptive strategies under which the node pair sampled next depends on the previously sampled pairs, and the corresponding outcomes. For both classes of sampling strategy, we identify fundamental performance limits satisfied by {\it any} joint sampling and clustering algorithm, and also devise simple algorithms that approach these limits. These results allow to quantify the gain achieved using adaptive sampling, and to determine how the observation budget has to scale with the number of users so as to ensure an asymptotically accurate community detection (the proportion of misclassified users tends to 0 as the number of users grows large).

\medskip
\noindent
{\bf Contributions.} We consider networks consisting of $n$ users or nodes with non-overlapping communities, and inspired by the celebrated stochastic block model, see \cite{holland1983}, we assume that the outcome of a node pair observation is positive (equal to 1) with probability $p$ if the nodes belong to the same community, and with probability $q<p$ otherwise. $p$ and $q$ may depend on the network size  $n$. We make no assumptions on $p$ and $q$. In particular, our results cover both the case of {\it dense} interactions where $p,q=\Theta(1)$ as $n$ grows large, and the case of {\it sparse} interactions where $p,q=o(1)$ as $n$ grows large. The observation budget is denoted by $T$, and typically depends on $n$ as well.

\medskip
\noindent
{\it a. Fundamental limits.} For any set of parameters $p$ and $q$, we derive asymptotic lower bounds on the expected proportion of misclassified nodes $\mathbb{E}[\varepsilon^\pi(n,T)]$ satisfied by any clustering algorithm $\pi$ in the case of non-adaptive random sampling strategies and by any joint sampling strategy and clustering algorithm $\pi$ in the case of adaptive sampling. We also give necessary conditions on $T$, $n$, $p$, and $q$ for asymptotically accurate community detection. More precisely:
\begin{itemize}
\item {\it Non-adaptive sampling.} Under any non-adaptive random sampling strategy, if there exists an asymptotically accurate clustering algorithm $\pi$, in the sense that $\lim_{n\to\infty}\mathbb{E}[\varepsilon^\pi(n,T)]=0$, then\footnote{$KL(p,q)=p\log(p/q)+(1-p)\log((1-p)/(1-q))$.} 
:
\begin{equation}\label{eq:c1}
{T\over n}=\omega(1),\quad\hbox{and}\quad {T\over n}\min(KL(q,p),KL(p,q))=\omega(1).
\end{equation}
\item {\it Adaptive sampling.} If there exists an asymptotically accurate joint adaptive sampling strategy and clustering algorithm $\pi$, then:
\begin{equation}\label{eq:c2}
\min\{1-q, p \}{T\over n}=\Omega(1),\quad\hbox{and}\quad {T\over n}\max(KL(q,p),KL(p,q))=\omega(1).
\end{equation}
\end{itemize}
The gain achieved using adaptive sampling can be significant when for example, $q$ is much smaller than $p$. To derive our performance bounds, we leverage change-of-measure arguments similar to those used in bandit optimization to provide regret lower bounds. This contrasts with most techniques used in statistical inference to obtain such bounds  (most often there, the analysis relies on Fano's inequality).

\medskip
\noindent
{\it b. Cluster Algorithms for Non-adaptive Sampling.} For non-adaptive random sampling strategies, we devise a low-complexity clustering algorithm, referred to as Spectral Partition (SP). The algorithm first constructs an {\it observation} matrix where the outcomes of the node pair observations are reported. After appropriate trimming, the spectral properties of the matrix (the largest eigenvalues and the corresponding eigenvectors) are exploited to derive rough estimates of the communities. These estimates are then improved using a recursive greedy procedure inspired by the $k$-mean algorithm. We prove that the SP algorithm is asymptotically accurate under conditions (\ref{eq:c1}), i.e., it is order-optimal. This implies in particular that the necessary conditions (\ref{eq:c1}) for asymptotically accurate detection are tight (they are also sufficient). We further analyse the performance of the SP algorithm. For example for networks with two communities of respective sizes $\alpha n$ and $(1-\alpha)n$, using techniques from random matrix theory, we prove that under conditions (\ref{eq:c1}), $\varepsilon^{SP}(n,T)\le \exp(-{(p-q)^2\over 20 p}{\alpha_1 T\over n})$ with high probability\footnote{An event $\chi$ occurs with high probability, if $\lim_{n \to \infty} \mathbb{P}[\chi] = 1$.}, where $\alpha_1 n $ is  the size of the smallest cluster. 

\medskip
\noindent
{\it c. Joint Adaptive Sampling and Clustering Algorithms.} We also propose a joint sampling and clustering algorithm, referred to as Adaptive Spectral Partition (ASP). The algorithm exploits the idea of {\it spatial coupling} recently advocated in compressed sensing by \cite{krzakala2012} and coding theory by \cite{kudekar2011}. More precisely, under ASP, we first use a positive fraction of the observation budget to classify a small proportion of nodes with very high accuracy. To do so we use the SP algorithm. After this first step, we obtain subsets of the communities, referred to as reference kernels, and for which we have strong probabilistic guarantees. The remaining observation budget is then used to attach the remaining nodes to the various reference kernels in an adaptive way. We establish that the ASP algorithm is asymptotically accurate under conditions (\ref{eq:c2}), which implies that (\ref{eq:c2}) are necessary and sufficient conditions for asymptotically accurate detection. The performance analysis of the ASP algorithm reveals for example that for networks with two communities, under conditions (\ref{eq:c2}), $\varepsilon^{ASP}(n,T)\le \exp(-{\alpha_1 T\over 3Kn}(KL(q,p)+KL(p,q)))$ with high probability. We compare the performance of SP and ASP using numerical experiments, and confirm that adaptive sampling may yield significant performance gains.

\medskip
\noindent
{\bf Related work.} Community detection has attracted a lot of attention in different scientific fields recently, and the topic is too vast for a detailed review of existing results here. \cite{newman2013}, \cite{coja2010}, \cite{mossel2013} and references therein cover a large part of the literature, from physics, computer science, and mathematics perspectives. As already mentioned, the starting point of most of the studies is an observed graph of interaction. In such as case, detecting communities boils down to a graph partitioning problem, that one can solve using spectral methods \cite{boppana1987eigenvalues}, \cite{mcsherry2001spectral}, \cite{dasgupta2006spectral}, \cite{chaudhuri2012spectral}, compressed sensing and matrix completion ideas \cite{chen2012}, \cite{chatterjee2012}, or other techniques \cite{Jerrum1998155}. Our model and approach are different: we address the problem of gathering information on node interactions, and that of identifying communities jointly. As far as we know, we provide the first set of results for this framework. 

The stochastic block model \cite{holland1983}, also referred to as the planted partition model, has been extensively used to assess the performance of community detection algorithms, see e.g. \cite{rohe2011}, \cite{decelle2011}. Our model is much more general, and covers as a particular case the stochastic block model (the latter corresponds to the case of non-adaptive sampling strategy where one has one observation per node pair, i.e., $T=n(n-1)/2$). There is a rich literature on community detection for the stochastic block model. In the dense regime, where $p,q=\Theta(1)$, most previous work focuses on identifying conditions under which a given algorithm recovers the clusters exactly, see e.g. \cite{mcsherry2001spectral}, \cite{condon2001}. For example, in \cite{chen2012} the authors show that communities can be extracted if $p-q\ge \Omega(\sqrt{p/n}\log(n)^2)$. In the sparse regime where $p,q=o(1)$, the main focus recently has been on identifying the phase transition threshold (a condition on $p$ and $q$) for reconstruction. It was conjectured in \cite{decelle2011} that if $p-q<\sqrt{2n(p+q)}$ (i.e., under the threshold), no algorithm can perform better than a simple random assignment of users to clusters, and above the threshold, clusters can partially be recovered. The conjecture was recently proved in \cite{mossel2012stochastic}, \cite{massoulie2013}, \cite{mossel2013}. A good survey of other existing results for the sparse regime can be found in \cite{coja2010}, and \cite{chen2012}. In this paper, we provide a unified (in dense and sparse regimes) treatment of the stochastic block model, and derive, as far as we know, the first necessary and sufficient conditions for asymptotically accurate community detection valid under any set of parameters $p$ and $q$. Necessary conditions for accurate detection are not derived in the aforementioned work. 

This paper covers more than the stochastic block model. It provides a systematic analysis of joint sampling strategies and clustering algorithms. For example, from our results, we can quantify the number of observations required to accurately detect communities when under the stochastic block model, this is not possible (i.e. when we are under the phase transition threshold for reconstruction).

\section{Models and Objectives}\label{sec:model}

We consider a network consisting of a set $V$ of $n$ nodes. $V$ admits a hidden partition of $K$ non-overlapping subsets $V_1,\ldots,V_K$ ($V=\bigcup_{k=1}^KV_k$). The size of class or cluster $V_k$ is $\alpha_k\times n$ for some $\alpha_k>0$. Without loss of generality, we assume that $\alpha_1 \le \alpha_2 \le \dots \le \alpha_K$. We assume that when the network size $n$ grows large, the number of communities $K$ and their relative sizes are kept fixed. By observing random pairwise interactions between nodes, we wish to recover the hidden partition. Let $E=V\times V$ be the set of node pairs. Pairs of nodes are successively sampled or observed. When a pair of nodes is sampled, these nodes are more likely to interact if they belong to the same community. More precisely, nodes of the same community interact with probability $p$, and nodes of different communities interact with probability $q$, with $q<p$. If for the $t$-th observation, node pair $(v,w)$ is sampled, the outcome $X_{vw}(t)$ is 1 if nodes interact, in which case we say that the observation is positive, and 0 otherwise. The Bernoulli random variables $X_{vw}(t)$'s are independent across nodes pairs $(v,w)$ and time $t$. We have a budget of $T$ observations, and $T$ can be either smaller than or equal to $n(n-1)/2$, in which case we say that the network is under-sampled, or larger than $n(n-1)/2$, in which case the network is over-sampled. We are primarily interested in large networks, and wish to design algorithms able to recover the partition accurately when $n$ is large. Naturally, the network parameters $p$ and $q$, as well as the observation budget $T$, may depend on $n$. 

\subsection{Sampling strategies}

We consider different types of sampling strategies.

\noindent
{\bf Random Sampling.} Here the sequence of observed node pairs is random, and we are mainly interested in two types of such sequences:\\
(1) Uniform Random Sampling (URS-1): for the $t$-th observation, the observed pair of nodes is chosen uniformly at random. \\
(2) Uniform Random Sampling without Replacement (URS-2): Assume that the budget $T=mn(n-1)/2+r$ where $m\in \mathbb{N}$ and $r\in\{0,\ldots,n(n-1)/2-1\}$. Here each pair $(v,w)$ is first observed $m$ times, and for the $r$ remaining observations, node pairs are selected uniformly at random without replacement (each pair is observed at most $m+1$ times). 

\noindent
{\bf Adaptive Sampling.} It may be more efficient to design the sequence of observed node pairs in an adaptive manner. We could select the node pair to be observed next depending on the past observations. In this case, the $(t+1)$-th observed pair, or more generally its distribution (in case of randomized sampling strategy), depends on $(e_s,X_s, s=1,\ldots, t)$, where $e_s$ denotes the $s$-th observed node pair, and $X_s$ is the corresponding interaction outcome. 

Under all sampling strategies, after the $T$ observations, one applies a clustering algorithm to recover the initially hidden partition. Such an algorithm $\pi\in \Pi$ maps the observations $(e_t,X_t, t=1,\ldots, T)\in (E\times \{0,1\})^T$ to an estimated partition $(\hat{V}_1,\ldots,\hat{V}_k)$ of the set $V$. The performance of the joint sampling strategy and clustering algorithm $\pi$ is quantified using the proportion $\varepsilon^\pi(n,T)$ of nodes that are misclassified. We say that $\pi$ is asymptotically accurate when $\lim_{n\to\infty} \mathbb{E}[\varepsilon^\pi(n,T)]=0$ (note that in the previous limit, $T$, $p$, and $q$ typically vary with $n$). For a given sampling strategy, we are interested in deriving conditions on $n$, $T$, $p$, and $q$, such that there exists an asymptotically accurate algorithm $\pi\in \Pi$.

\subsection{Stochastic Block Model with Labels}

To study the performance of non-adaptive sampling strategies in both under and over-sampling scenarios, it is instrumental to introduce the so-called {\it Stochastic Block Model with Labels} (SBML), see \cite{heimlicher2012community}. In SBML, the outcome of an observation is a label $\ell\in {\cal L}$, and each node pair is observed once -- we have exactly $n(n-1)/2$ observations. The observation of a node pair $e$ yields a label $\ell(e)$ equal to $\ell$ with probability $p(\ell)$ if the two nodes are within the same community, and with probability $q(\ell)$ otherwise. In the SBML, one may think of a label $\ell$ as a type of interaction between two nodes. In what follows, we denote by $0\in {\cal L}$ a particular label. The latter typically represents the absence of interaction between two nodes.

In the SBML, one has access to the sampled labels of each node pair, and one applies a clustering algorithm to retrieve the communities. Such an algorithm $\pi\in \Pi'$ maps the observations $(\ell(e), e\in E)$ to an estimated partition $(\hat{V}_1,\ldots,\hat{V}_K)$ of the set $V$.

Non-adaptive sampling strategies can be represented as particular examples of the SBML. Indeed, the label of a node pair can represent all the information gathered on this pair using the $T$ observations. We provide below a way of representing URS-1 and under-sampled URS-2 sampling strategies using the SBML. The over-sampled URS-2 sampling strategies can be mapped to the SBML similarly.

\begin{example} (URS-1) Let $\beta=2/(n(n-1))$. The set of labels is ${\cal L}=\{(m,z),m\in \{0,\ldots T\}, z\in\{0,\ldots,m\}\}$. After the $T$ observations, a node pair has the label $(m,z)$ if this pair has been observed $m$ times, and that the interaction outcomes have been equal to 1 $z$ times. For example, a pair has label  $(0,0)$ if it has not been observed. The label distribution is then defined as: for all $(m,z)\in {\cal L}$, $p(m,z)={T\choose m}\beta^m(1-\beta)^{T-m} {m\choose z}p^z(1-p)^{m-z}$, and $q(m,z)={T\choose m}\beta^m(1-\beta)^{T-m} {m\choose z}q^z(1-q)^{m-z}.$
\end{example}

\begin{example} (Under-sampled URS-2) To represent random sampling strategies without replacement in the under-sampled scenario using the SBML, we introduce three labels $\emptyset$, 0, and 1, i.e., ${\cal L}=\{\emptyset, 0, 1\}$. A node pair has label $\emptyset$ if it has not been observed, 0 if it has been observed (once) and if the outcome is 0, and 1 if it has been observed and if the outcome is 1. Let $\beta=2T/(n(n-1))$ denote the proportion of observed node pairs. The label distribution is: $p(\emptyset)=q(\emptyset)=1-\beta$, $p(0)=(1-p)\beta$, $q(0)=(1-q)\beta$, $p(1)=p\beta$, and $q(1)=q\beta$. 
\end{example}


\section{Lower Bounds}

In this section, we derive lower bounds of the average proportion of misclassified nodes under the various types of joint sampling strategy and clustering algorithm. We provide a lower bound first for the SBML and non-adaptive sampling strategies, and then for adaptive sampling strategies. The lower bounds allow us to identify a necessary condition for asymptotically accurate community detection.

\subsection{Non-adaptive Random Sampling}

\subsubsection{The SBML}

We denote by $\varepsilon^\pi(n)$ the proportion of misclassified nodes under a given clustering algorithm $\pi\in \Pi'$. Again we say that a clustering algorithm $\pi\in \Pi'$ is asymptotically accurate if $\lim_{n\to\infty} \mathbb{E}[\varepsilon^\pi(n)]=0$.  The following theorem provides a lower bound of the expected proportion of misclassified nodes satisfied by any asymptotically accurate algorithm. Recall that $\alpha_1$ defines the size of the smallest cluster (i.e., the latter is of size $\alpha_1 n$).

\begin{theorem}
In the sparse regime ($\lim_{n\to \infty}\frac{\sum_{\ell\neq 0}p(\ell)+q(\ell)}{\min\{p(0),q(0)\}}=0$), for any asymptotically accurate algorithm $\pi\in \Pi'$, we have:
$$\lim\inf_{n \rightarrow \infty} {4\mathbb{E}[\varepsilon^\pi(n)]\over \alpha_1\exp(-4(\alpha_1 + \alpha_2)\tau(n))} \ge 1 ,\quad \mbox{where}\quad \tau(n) = \sum_{\ell \in {\cal L} } n\frac{(p(\ell)-q(\ell))^2}{p(\ell)+q(\ell)} .$$

\label{thm:lower-bound}
\end{theorem}


\subsubsection{URS-1 and URS-2 Sampling Strategies}

Theorem \ref{thm:lower-bound} can be applied to the various aforementioned non-adaptive sampling strategies. Actually, for the strategies considered here, the results presented in Theorem \ref{thm:lower-bound} can be improved: we derive a universal non-asymptotic lower bound on the average proportion of misclassified nodes valid under random sampling strategies URS-1 and URS-2, and for all set of parameters $n$, $T$, $p$, and $q$. Recall that $\alpha_2$ defines the size of the second smallest cluster (i.e., the latter is of size $\alpha_2 n$). 

\begin{theorem}\label{thm:l1}
Under URS-1 and URS-2 sampling strategies, for any clustering algorithm $\pi\in\Pi$, we have: for all $T$, $p$, $q$, and $n$,
\begin{equation}\label{eq:nonas}
{\mathbb{E}[\varepsilon^\pi(n,T)]\ge {\alpha_1 \over 4} \exp(-\kappa_1(n,T))},
\end{equation}
where
\begin{align*}
\kappa_1(n,T)=& T{2(\alpha_1+\alpha_2)\over n} \min \{ KL(q,p), KL(p,q)\}  \cr & ~+2\sqrt{\frac{4T(\alpha_1 + \alpha_2)}{n}\left[\min\{q,1-p\} \left(
    \log\frac{p(1-q)}{q(1-p)}\right)^2  + \left(
    \log(\min\{\frac{p}{q},\frac{1-q}{1-p} \})\right)^2 \right]}.
\end{align*}
As a consequence, for any asymptotically accurate clustering algorithm $\pi\in\Pi$ (i.e., satisfying $\lim_{n\to\infty} \mathbb{E}[\varepsilon^\pi(n,T)]=0$), we have:
\begin{equation}\label{eq:univ2}
{T\over n}=\omega(1),\quad {T\over n}\min(KL(q,p),KL(p,q))=\omega(1),
\end{equation}
\begin{equation}\label{eq:univ}
\hbox{and}\quad \lim\inf_{n\to\infty} {4\mathbb{E}[\varepsilon^\pi(n,T)]\over \alpha_1 \exp(-{2(\alpha_1 + \alpha_2)T\over n}\min(KL(q,p),KL(p,q)))}\ge 1.
\end{equation}
\end{theorem}

(\ref{eq:univ2}) provides two necessary conditions for asymptotically accurate community detection. We show in the next section that these conditions are also sufficient, i.e., we propose a clustering algorithm that is asymptotically accurate when ${T\over n}=\Omega(1)$ and ${T\over n}\min(KL(q,p),KL(p,q))=\omega(1)$. Note that the results of Theorem \ref{thm:l1} hold for arbitrary parameters $p$ and $q$. 

For {\it dense} interactions where $p,q=\Theta(1)$, we need $T(p-q)^2/n=\omega(1)$\footnote{We repeatedly use the facts that for $q\le p$, $2(p-q)^2\le KL(q,p)\le (p-q)^2/(p(1-p))$, and $KL(q,p)\sim (p-q)^2/(p(1-p))$ as $q\to p^-$.} to get an asymptotically accurate detection. This is in agreement with existing results for the stochastic block model (URS-2 sampling strategy with $T=n(n-1)/2$), see Table 1 in \cite{chen2012}: the best known algorithms recover the communities exactly ($n\varepsilon(n,T) = 0$) with high probability when $p-q=\Omega(\sqrt{\log(n)/n})$, and our lower bound says that to obtain $\mathbb{E}[n \varepsilon (n,T)] < 1$, we need $p-q=\Omega(\sqrt{\log(n)/n})$. 

For {\it sparse} interactions where $p,q=o(1)$, we need $T(p-q)^2/(pn)=\omega(1)$ to get an asymptotically accurate detection. For example, when $p=a/n$ and $q=b/n$ for some constants $a>b$, then we need ${T\over n^2}=\omega(1)$ for accurate detection. Note that in this case, for the classical stochastic block model, i.e., for $T=n(n-1)/2$, a necessary and sufficient condition to be able to devise an algorithm that performs better than assigning nodes randomly to communities is $(a-b)>\sqrt{2(a+b)}$ \cite{mossel2012stochastic}, \cite{massoulie2013}. Our result indicates that when targeting an asymptotically accurate detection, we need much more observations (e.g. $T=\log\log(n)n^2$).

It should be finally observed that the results of Theorem \ref{thm:l1} do not depend on the way node pairs are sampled, provided that they are randomly selected. In particular, we do not expect that sampling without replacement outperforms purely random sampling (with equal observation budget). 

\subsection{Adaptive Sampling}

Next we derive similar asymptotic lower bounds on the expected proportion of misclassified nodes in the case of adaptive sampling strategies. The proof of the following theorem is more involved than that of Theorem \ref{thm:l1}; it relies on a change-of-measure argument and on Doob's maximal inequality. 

\begin{theorem}\label{thm:la}
For any asymptotically accurate joint adaptive sampling strategy and clustering algorithm $\pi\in\Pi$ (i.e., satisfying $\lim_{n\to\infty} \mathbb{E}[\varepsilon^\pi(n,T)]=0$), we have: 
\begin{equation}\label{eq:univ3}
\min\{p,1-q\}{T\over n}=\Omega(1)  \quad\hbox{and}\quad {T\over n}\max(KL(q,p),KL(p,q))=\omega(1).
\end{equation}
In addition, when $\frac{-\log \mathbb{E}[\varepsilon^\pi(n,T)]}{ \max\{\log \frac{p}{q},\log \frac{1-q}{1-p}\} } = \omega(1)$, the following holds:
\begin{equation}\label{eq:univ4}
 \lim\inf_{n\to\infty} {\mathbb{E}[\varepsilon^\pi(n,T)]\over \exp(-{8T\over \min\{1/2, 1-\alpha_K\} n}\max(KL(q,p),KL(p,q)))}\ge 1,
\end{equation}
where $\alpha_k$ defines the size of the largest cluster (i.e., the latter is of size $\alpha_K n$).
\end{theorem}

In view of Theorems \ref{thm:l1} and \ref{thm:la}, adaptive sampling is expected to outperform random sampling (with equal observation budget) when $\min(KL(q,p),KL(p,q))\ll \max(KL(q,p),KL(p,q))$. For example, in the case of sparse interactions, if $q=p^\gamma$ with $\gamma>1$, then the necessary conditions for asymptotically accurate detection reduce to ${pT\over n}=\omega(1)$ and ${p\log(1/p)T\over n}=\omega(1)$ under non-adaptive random sampling and adaptive sampling, respectively. Note also that even if the necessary conditions for accurate detection are identical under any sampling strategy (non-adaptive or adaptive), then the lower bound on the expected proportion of misclassified nodes is improved under adaptive sampling. In the next section, we show that the necessary conditions (\ref{eq:univ3}) for accurate detection are sufficient, i.e., we propose a clustering algorithm that is asymptotically accurate when $\min\{p,1-q \}{T\over n}=\Omega(1)$ and ${T\over n}\max(KL(q,p),KL(p,q))=\omega(1)$. 

\section{Algorithms}

In this section, we present simple clustering algorithms for non-adaptive sampling strategies, as well as joint adaptive sampling and clustering algorithms. We provide upper bounds on the proportions of misclassified nodes under these algorithms, and establish that they are order-optimal: they are asymptotically accurate as soon as conditions (\ref{eq:c1}) or (\ref{eq:c2}) are satisfied.

\subsection{Non-Adaptive Sampling}

We first propose Spectral Partition (SP), a clustering algorithm for non-adaptive URS-1 or URS-2 sampling strategies. From the $T$ observations, we construct a matrix $A\in\mathbb{N}^{n\times n}$: for any pair $(v,w)$, $A_{vw}$ is equal to the number of positive observations of node pair $(v,w)$. In particular, if $(v,w)$ has not been observed, $A_{vw}=0$. Note that for all pair $(v,w)$, $\mathbb{E}[A_{vw}]={2T\over n(n-1)}p$ if $v$ and $w$ are in the same cluster, and 
$\mathbb{E}[A_{vw}]={2T\over n(n-1)}q$ otherwise (the expectation is taken accounting for the randomness in both the number of times node pair $(v,w)$ is observed, and the corresponding outcomes). Matrix $\mathbb{E}[A]$ is symmetric and of rank $K$, and its eigenvectors identify the clusters. For example, if $K=2$, the two eigenvalues of $\mathbb{E}[A]$ are $\overline{\lambda}_1={T\over n-1}(p+\sqrt{p^2-4\alpha(1-\alpha)(p^2-q^2)})$ and $\overline{\lambda}_2={T\over n-1}(p-\sqrt{p^2-4\alpha(1-\alpha)(p^2-q^2)})$, respectively, where $\alpha=\alpha_1$ is the proportion of nodes in the first cluster. Assume now that the first cluster corresponds to nodes $1,\ldots,\alpha n$, then the eigenvectors of $\mathbb{E}[A]$ are $(1,\ldots,1,a_i,\ldots,a_i)$ where the first $\alpha n$ components are equal to 1, and $a_i={\overline{\lambda}_i{n-1\over 2T}-p\alpha\over (1-\alpha)q}$, for $i=1,2$.  

From a spectral analysis of $A$, we expect to accurately recover the clusters if $\frac{(p-q)^2}{p+q}\frac{T}{n} \gg 1$. Indeed it can be seen that the eigenvalues of $\mathbb{E}[A]$ are $\Omega((p-q){T\over n})$. In addition, the noise matrix $X=A-\mathbb{E}[A]$ satisfies $\| X \| =O (\sqrt{\frac{T}{n}(p+q)})$ provided that the number of observations per node pair does not exceed $\log(n)$ (this is a simple consequence of random matrix theory, see e.g. \cite{tao2012, chatterjee2012}). The SP algorithm whose pseudo-code (Algorithm 1) is presented below, exploits this observation and may be seen as an extension of algorithms proposed in \cite{coja2010} to recover clusters in the simple stochastic block model ($T=n(n-1)/2$ and URS-2 sampling strategy). Our algorithm works for any observation budget and any random sampling strategy, and its performance analysis is much simpler than that presented in \cite{coja2010}.

\begin{algorithm}[t!]
   \caption{Spectral Partition}
   \label{alg:partition}
\begin{algorithmic}
\STATE {\bfseries Input:} Observation matrix $A$.
\STATE  {\bf 1. Trimming.} Construct $A_{\Gamma}=(A_{vw})_{v,w\in \Gamma}$ where $\Gamma = \{v:
  \sum_{w \in V} A_{vw} \le 5K \frac{\sum_{(v,w) \in E } A_{vw}}{n} \}$.
\STATE {\bf 2. Spectral Decomposition.} Run Algorithm 2 ($K=2$) or Algorithm 3 ($K\ge 3$)
\STATE with input $A_{\Gamma}, \frac{ \sum_{(v,w) \in E } A_{vw} }{n^2}$, and output $(S_k)_{k=1,\ldots,K}$.
\STATE {\bf 3. Improvement.}
   \STATE $S^{(0)}_k \leftarrow S_k ,$ for all $k$
   \FOR{$i=1$ {\bfseries to} $\log n$ }
     \STATE $S^{(i)}_k \leftarrow \emptyset ,$ for all $k$
      \FOR{$v \in V$}
  \STATE Find $k^{\star} = \arg \max_{k} \{\sum_{w \in V} A_{vw} / |
  S^{(i-1)}_k | \} $ (tie broken uniformly at random)
   \STATE $S^{(i)}_{k^{\star}} \leftarrow S^{(i)}_{k^{\star}} \cup \{ v \}$
   \ENDFOR
   \ENDFOR
\STATE $\hat{V}_k \leftarrow S_k^{(i)}$, for all $k$
   \STATE {\bfseries Output:} $(\hat{V}_k)_{k=1,\ldots,K}$.
\end{algorithmic}
\end{algorithm}

The algorithm has three steps. \\
{\bf 1. Trimming.} We first trim the observation matrix $A$, i.e., we keep the entries corresponding to a set $\Gamma$ of nodes that did not get too many positive observations. More precisely, $\Gamma = \{v: \sum_{w \in V} A_{vw} \le 10 \frac{\sum_{(v,w) \in E } A_{vw}}{n} \} $. The resulting trimmed observation matrix is denoted by $A_\Gamma$. \\
{\bf 2. Spectral decomposition.} We then extract the clusters from the spectral analysis of $A_\Gamma$. We present a simple method (Algorithm 2) when $K=2$, exploiting the fact that clusters can be recovered just looking at the signs of the components of the eigenvectors corresponding to the two largest eigenvalues of $A_\Gamma$. When $K\ge 3$, we extract the clusters from the column vectors of the rank-$K$ approximation matrix $\hat{A}$ of $A_\Gamma$. This rank-$K$ approximation is obtained by singular value decomposition and by keeping the $K$ largest singular values and the corresponding eigenvectors, see \cite{chatterjee2012}. Our algorithm exploits the fact that the column vectors corresponding to nodes in the same clusters should be relatively close to each other. We use the distance between these vectors to classify nodes, in the spirit of the $k$-means clustering algorithm. In the pseudo-code, $\hat{A}_v$ denotes the column vector of $\hat{A}$ corresponding to node $v$, and $\|\cdot\|$ refers to the euclidian distance.\\
{\bf 3. Improvement.} Finally, we further improve the results. After the spectral decomposition step, the identified clusters $(S_k)_{k=1,\ldots,K}$ are good approximations of the true clusters. The improvement is obtained by sequentially considering each node and by moving the node to the cluster with which it has the largest number of positive observations.

\begin{algorithm}[t!]
   \caption{Spectral decomposition (for $K=2$)}
   \label{alg:trimspec}
\begin{algorithmic}
\STATE {\bfseries Input:} $A_{\Gamma}$, $\Gamma$.
\STATE $x_1$ and $x_2 ~ \leftarrow$ the eigen vectors of $A_{\Gamma}$ corresponding to the two largest eigenvalues.
\IF{$\big( \sum_{v \in \Gamma} x_1 (v) \big)\cdot \big(\sum_{v \in
    \Gamma} x_1 (v)\big) >0$}\STATE{$x_2 \leftarrow -x_2$}
\ENDIF
\STATE $\hat{x} \leftarrow x_1 + x_2 -\frac{1}{| \Gamma |} {\bm J}(x_1 +
  x_2),$ where $\bm{J}$ denotes the $\Gamma \times \Gamma$ matrix
  filled with $1.$
\STATE $S_1 \leftarrow \{v \in \Gamma : \hat{x}(v) >0 \}$ and $S_2 \leftarrow \{v \in \Gamma : \hat{x}(v) <0 \}$
\STATE For all $v \notin S_1 \cup S_2,$ randomly place $V$ in
$S_1$ or $S_2$
\STATE {\bfseries Output:} $(S_1 , S_2)$.
\end{algorithmic}
\end{algorithm}

\begin{algorithm}[t!]
   \caption{Spectral decomposition (for $K\ge 3$)}
   \label{alg:specg}
\begin{algorithmic}
   \STATE {\bfseries Input:} $A_{\Gamma}, \frac{ \sum_{(v,w) \in E } A_{vw} }{n^2}$ 
\STATE $\hat{A} \leftarrow $ $K$-rank approximation of $A_{\Gamma}$
\FOR{$i=1$ {\bfseries to} $\log n$ }
\STATE $Q_{i,v} \leftarrow \{ w \in {V} :
\| \hat{A}_w  -\hat{A}_v\|^2 \le i \frac{\sum_{(v,w) \in E } A_{vw} }{100n^2} \} $
\STATE $T_{i,0}\leftarrow \emptyset$
\FOR{$k=1$ {\bfseries to} $K$ }
\STATE $v_k^{\star} \leftarrow \arg \max_{v} | Q_{i,v}\setminus \bigcup_{l=1}^{k-1} T_{i,l} |$ 
\STATE $T_{i,k} \leftarrow Q_{i,v_k^{\star}} \setminus \bigcup_{l=1}^{k-1} T_{i,l} $ and $\xi_{i,k} \leftarrow \sum_{v \in T_{i,k}}  \hat{A}_v/ |T_{i,k}| .$
\ENDFOR
\FOR{$v \in V \setminus ( \bigcup_{k=1}^K T_{i,k} )$}
\STATE $k^{\star} \leftarrow \arg \min_{k} \| \hat{A}_v -\xi_{i,k}\|$ 
\STATE $T_{i,k^{\star}} \leftarrow T_{i,k^{\star}} \cup \{v\}$
\ENDFOR
\STATE $r_i  \leftarrow \sum_{k=1}^K \sum_{v \in T_{i,k}} \| \hat{A}_v -\xi_{i,k}\|^2$
\ENDFOR

\STATE $i^{\star} \leftarrow \arg \min_{i} r_i.$
\STATE $S_k\leftarrow T_{i^\star,k}$ for all $k$
\STATE {\bfseries Output:} $(S_k)_{k=1,\ldots,K}$.
\end{algorithmic}
\end{algorithm}

In the following theorem, we analyze the performance of the first two steps of the SP algorithm (we stop after the Spectral Decomposition step, and do not apply the improvement step). 

\begin{theorem} Assume that $\frac{(p-q)^2}{p}{\alpha_1 T\over n} =\omega(1)$. Under URS-1 or URS-2 sampling strategy with $T$ observations, after Step 2 (Spectral decomposition) in the Spectral Partition algorithm, there exists a permutation $\sigma$ of $\{1,\ldots,K\}$ such that:
\begin{equation*}
\lim_{n\to\infty} \mathbb{P}\left[\frac{1}{n}|\bigcup_{k=1}^K V_{\sigma(k)} \setminus S_k| = 0\right]=1.
\end{equation*} \label{thm:specdecom}
\end{theorem}

Most often $p$ and $q$ are such that $\frac{p-q}{p}$ does not tend to 0, in which case we say that $p$ and $q$ are generic. For generic $p$ and $q$, when the necessary condition for accurate detection (\ref{eq:c1}) is satisfied, one can easily check that $\frac{(p-q)^2}{p}{\alpha_1 T\over n} =\omega(1)$ (because $p\frac{T}{n} = \omega(1)$). In that case, the fraction of misclassified nodes goes to 0 after Step 2 of SP algorithm. We conclude that the combination of the Trimming and Spectral Decomposition steps in the SP algorithm is asymptotically accurate whenever an accurate detection is at all possible and $p$ and $q$ are generic. The next theorem provides an upper bound of the proportion of misclassified nodes under the SP algorithm.

\begin{theorem}\label{thm:algorithms} Assume that $\frac{(p-q)^2}{p}{\alpha_1 T\over n} =\omega(1)$ and $\frac{(p-q)^2}{20p}{\alpha_1 T\over n} \ge \log(p{T\over n})$. Under URS-1 or URS-2 sampling strategy with $T$ observations, the proportion of misclassified nodes under Spectral Partition satisfies, with high probability,
\begin{equation}\label{eq:up}
\varepsilon^{SP} (n,T) \le \exp \left(-\frac{(p-q)^2 }{20 p} \frac{\alpha_1  T}{n}\right).
\end{equation}
\end{theorem}

Again, for generic $p$ and $q$, when the necessary condition for accurate detection (\ref{eq:c1}) is satisfied, one can easily check that  $\frac{(p-q)^2}{20p}{\alpha_1 T\over n} \ge \log(p{T\over n})$ (because $\log(p\frac{T}{n}) \le 2\log((p-q)\frac{T}{n})$ and $(p-q)\frac{T}{n}=\omega(\log((p-q)\frac{T}{n}))$). In that case, (\ref{eq:up}) holds, and in particular, $\lim_{n\to\infty}\mathbb{E}[\varepsilon^{SP}(n,T)]=0$.

In rare cases, the necessary condition (\ref{eq:c1}) does not imply the conditions of Theorem~\ref{thm:specdecom} and Theorem~\ref{thm:algorithms}. For example, when both $p$ and $q$ tend to 1, $\min\{KL(p,q),KL(q,p)\}\frac{T}{n} = \omega(1)$ does not mean $\frac{(p-q)^2}{p}{\alpha_1 T\over n} =\omega(1)$  because $\min \{ KL(p,q) , KL(q,p) \} \ge \frac{1}{2}\frac{(p-q)^2}{2-p-q} = \omega(\frac{(p-q)^2}{p})$. Moreover, when $(p-q)\frac{T}{n}=\omega(\sqrt{p\frac{T}{n}})$ and $(p-q)\frac{T}{n} < \sqrt{20p\frac{T}{\alpha_1 n}\log(p\frac{T}{n})},$  $\frac{(p-q)^2}{20p}{\alpha_1 T\over n} < \log(p{T\over n})$.

\subsection{Adaptive Sampling}

Next we devise an adaptive sampling and clustering algorithm, referred to as Adaptive Spectral Partition (ASP),  that typically outperforms any algorithm with non-adaptive random sampling (it beats the lower bounds on $\mathbb{E}[\varepsilon(n,T)]$ obtained for random sampling). The adaptive algorithm is also order-optimal: ASP is asymptotically accurate under conditions (\ref{eq:c2}).

The method to sample node pairs and reconstruct clusters is inspired by the idea of {\it spatial coupling} recently used in coding theory \cite{kudekar2011}, and in compressed sensing \cite{krzakala2012}. For example, in compressed sensing, spatial coupling consists in identifying with very high accuracy a small proportion of the components of the unknown vector, and to propagate this accuracy to other components using their inherent correlations. Here, we first identify $K$ small subsets of nodes, referred to as {\it reference kernels}, and such that all nodes within the same kernel are very likely to belong to the same cluster, and nodes in different kernels are very likely in different clusters. We then grow the clusters starting from the references kernels. To get a very high accuracy on the reference kernels, we use a positive fraction of the observation budget to sample pairwise interactions within a small subset of nodes. The remaining budget is used to determine the cluster of the remaining nodes.

The algorithm has two main steps.\\
{\bf 1. Construction of the reference kernels.} Randomly select a set $S\subset V$ of cardinality $n/(5\log(n))$ (here we just need the $|S|$ scales as $n/\log(n)$), and apply the SP algorithm to $S$ using $T/5$ observations. This gives the reference kernels $(S_k)_{k=1,\ldots,K}$. In addition, during this first step, we derive $\hat{p}$ and $\hat{q}$, estimators of the probabilities $p$ and $q$ (these estimators are simply obtained by counting the observations whose outcome are equal to 1 intra- and inter-kernels). We expect to identify {\it good kernels} in the sense that: $(C1)$: there exists a permutation $\sigma$ of $\{1,\ldots,K\}$ such that $\forall k, |S_k \setminus {V}_{\sigma(k)} |=0$, and $(C2): \left|1- \frac{\hat{p}-\hat{q}}{p-q} \right| \le 10^{-2}$.

Note that in the first step, the observation budget per node is $\log (n)$ times larger than that we would have if SP was applied to $V$ using $T$ observations. Now from the performance analysis of SP, the fraction of misclassified nodes decreases exponentially with the budget per node. When $\frac{(p-q)^2 }{p+q }\frac{\alpha_1 T}{n} \ge C$ for some $C>1$, we get $\frac{(p-q)^2 }{p+q }\frac{\alpha_1 T \log(n)}{n} \ge C\log(n)$ if we change the budget from $T$ to $T\log(n)$. Thus in view of Theorem~\ref{thm:algorithms}, with high probability, $\varepsilon(n,T) \le 1/n^C < 1/ n$. Therefore, with high probability, the reference kernels have no error, and condition (C1) holds. (C2) also holds with high probability (a direct consequence of the law of large numbers).

{\bf 2. Classification of the remaining nodes.} In this second step, we classify the remaining nodes using the reference kernels. For each of these nodes, say node $v$, for all $k$, we sample the node pair $(v,w)$ for $w$ uniformly selected in $S_k$, and repeat this ${2T\over 3Kn}$ times. We record the number of positive observations $A_k$ between $v$ and kernel $S_k$. We assign $v$ to $S_k$ if for any $k'\neq k$, $A_k-A_{k'}\ge \gamma$ where the threshold $\gamma$ guarantees the quality of the assignment. We choose $\gamma=\frac{(\hat{p}-\hat{q})T}{2Kn}$. This choice is motivated by the observation that $\mathbb{E}[A_k - A_{k'}] \approx \frac{(\hat{p}-\hat{q})2T}{3Kn} $ when $v \in V_k$ and $k\neq k' $. This procedure is repeated until there is no remaining nodes or no remaining budget. The second step is adaptive since the number of times a particular node $v$ is tested depends on the previous observation outcomes.

The pseudo-code of ASP is presented below. The next theorem provides performance guarantees for the ASP algorithm.

\begin{algorithm}[t!]
   \caption{Adaptive Spectral Partition}
   \label{alg:adaptive}
\begin{algorithmic}
   \STATE {\bfseries Input:} Observation budget $T$. 
   \STATE {\bfseries 1. Initialization:} $\hat{V}_k = \emptyset$ for all $k$ and
   $R = V$ 
   \STATE {\bfseries 2. Find the reference kernels:} Build node set $S$ by randomly selecting $\frac{n}{5 \log n}$ nodes. 
\STATE Get $T/5$ random observations for pairs of nodes in $S$, and construct an observation matrix $A_S$.
   \STATE Run the Spectral Partition algorithm with input $A_S$, and output $(S_k)_{k=1,\ldots,K}$. $\hat{V}_k\leftarrow S_k$ for all $k$.
   \STATE {\bfseries 3. Estimate $p$ and $q$}
   \STATE $\hat{p} \leftarrow 
  \frac{\sum_{k=1}^{K}\sum_{(v,w) \in S_k \times S_k} A_{vw}}{\sum_{k} | S_k |^2}\frac{|S|^2}{2T} $ and $\hat{q} \leftarrow\frac{\sum_{k=1}^{K}\sum_{(v,w) \in S_k \times S^c_k} A_{vw}}{ |S|^2 - \sum_{k} | S_k
     |^2}\frac{|S|^2}{2T}$
\STATE {\bfseries 4. Classify the remaining nodes.}
\REPEAT
\STATE $R \leftarrow V\setminus (\bigcup_{k}\hat{V}_k   )$ 
\FOR{$v \in R$}
\STATE Randomly sample $\frac{2T}{3Kn}$ pairs between $v$ and $S_k$ for
all $k$
\STATE $A_{vw} \leftarrow$ number of positive observations for $(v,w)$ 
\STATE $k^{\star}(v) \leftarrow \arg\max_{k} \sum_{w \in S_k} A_{vw}$ 
\STATE $d^{\star}(v) \leftarrow \min_{k \neq k^{\star}} \sum_{w \in S_{k^{\star}}}
A_{vw} - \sum_{w' \in S_k} A_{vw'}$
\IF{$d^{\star}(v) \ge \frac{\hat{p}-\hat{q}}{2K}\frac{T}{n}$}
\STATE $\hat{V}_{k^{\star}(v)} \leftarrow \hat{V}_{k^{\star}(v)} \cup \{ v\}$
\ENDIF
\ENDFOR
\UNTIL{There exists no remaining node or budget}
\STATE If $R\neq\emptyset$, randomly assign $v \in R$ to $\{ \hat{V}_k \}$. Output $(\hat{V}_k)_{k=1,\ldots,K}$.
\end{algorithmic}
\end{algorithm}

\begin{theorem} When $\frac{(p-q)^2}{p+q}\frac{T}{n} =\Omega(1)$ and ${T\over n}\max(KL(q,p),KL(p,q))=\omega(1)$, the proportion of misclassified nodes under the Adaptive Spectral Partition algorithm satisfies, with high probability,
\begin{align}\label{eq:up2}
 \varepsilon^{ASP}(n,T) & \le \exp\left(-\frac{T}{3Kn} \big( KL(q,p)+KL(p,q) \big) \right).
\end{align}
\label{thm:adapt-results}
\end{theorem}

From the results of Theorem~\ref{thm:la}, for any adaptive sampling, to get accurate reconstruction, i.e., $\lim_{n\rightarrow \infty} \mathbb{E}[\varepsilon (n,T)] = 0$, the number of observations $T$ should satisfy ${T\over n}\max(KL(q,p),KL(p,q))=\omega(1)$ and $\min\{p,1-q\}{T\over n}=\Omega(1)$. This necessary condition implies $\frac{(p-q)^2}{p+q}\frac{T}{n} =\Omega(1)$ when $q$ does not tend to 1. Indeed, when $p$ does not go to 1, $\frac{(p-q)^2}{p+q}\frac{T}{n} = \omega(1)$  (because $KL(q,p) \le \frac{(p-q)^2}{p(1-p)}$ and $KL(p,q) \le \frac{(p-q)^2}{q(1-q)}$), and when $p$ tends to 1, $\frac{(p-q)^2}{p+q}\frac{T}{n} =\Omega(1)$ since $ \frac{(p-q)^2}{p(p+q)} = \Theta(1)$. Thus, when $q$ does not tend to 1, ASP is asymptotically accurate under the necessary condition (\ref{eq:c2}).



\section{Conclusion}\label{sec:conclusion}
In this paper, we studied the problem of community detection in networks using non-adaptive and adaptive sampling strategies. We derived necessary conditions under which an accurate detection is possible when the network size grows large, and presented algorithms that are accurate under these conditions. Our numerical experiments presented in appendix show that gathering information in an adaptive manner can significantly improve the detection accuracy.

\clearpage
\newpage

\bibliography{reference}

\appendix
\section{Numerical Experiments}

\begin{figure*}[t]
  \begin{center}
\subfigure[$p=10^{-3}$ and $q=p/20$]{
      \includegraphics[width = 0.31\columnwidth]{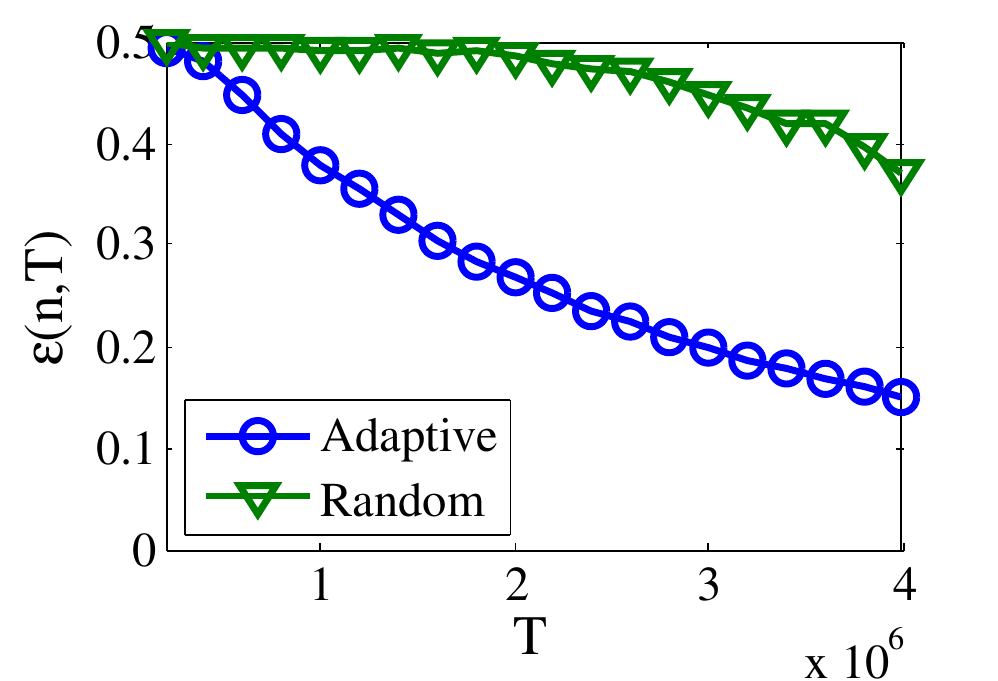}} \label{fig:p1}
\subfigure[$p=10^{-2}$ and $q=p/2$]{
\includegraphics[width = 0.31\columnwidth]{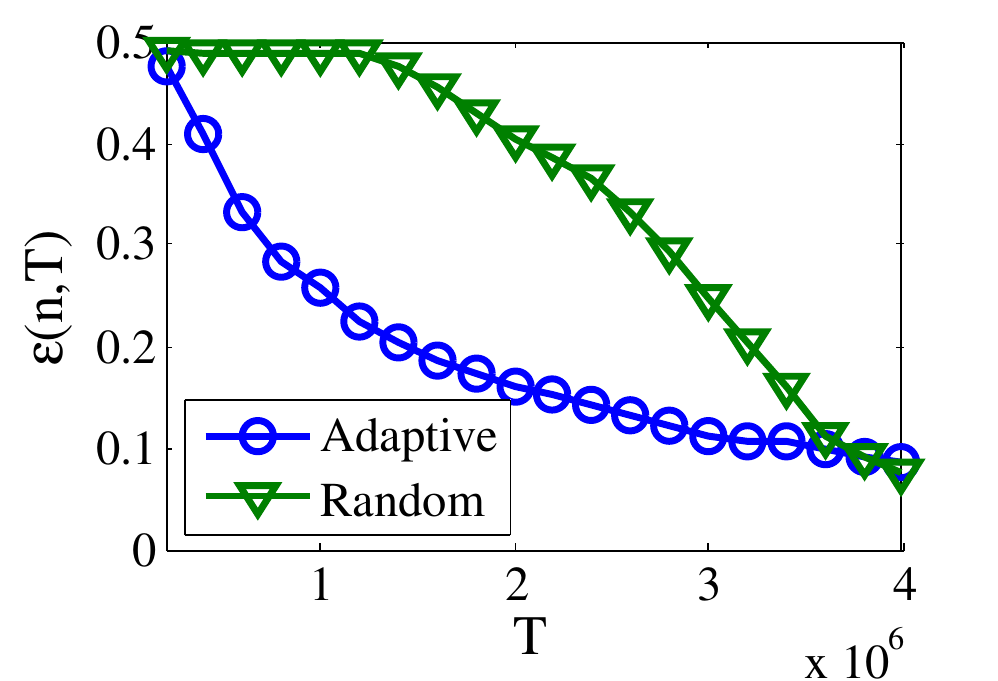} \label{fig:p2}
}
\subfigure[$p=10^{-1}$ and $q=p/2$]{
\includegraphics[width = 0.31\columnwidth]{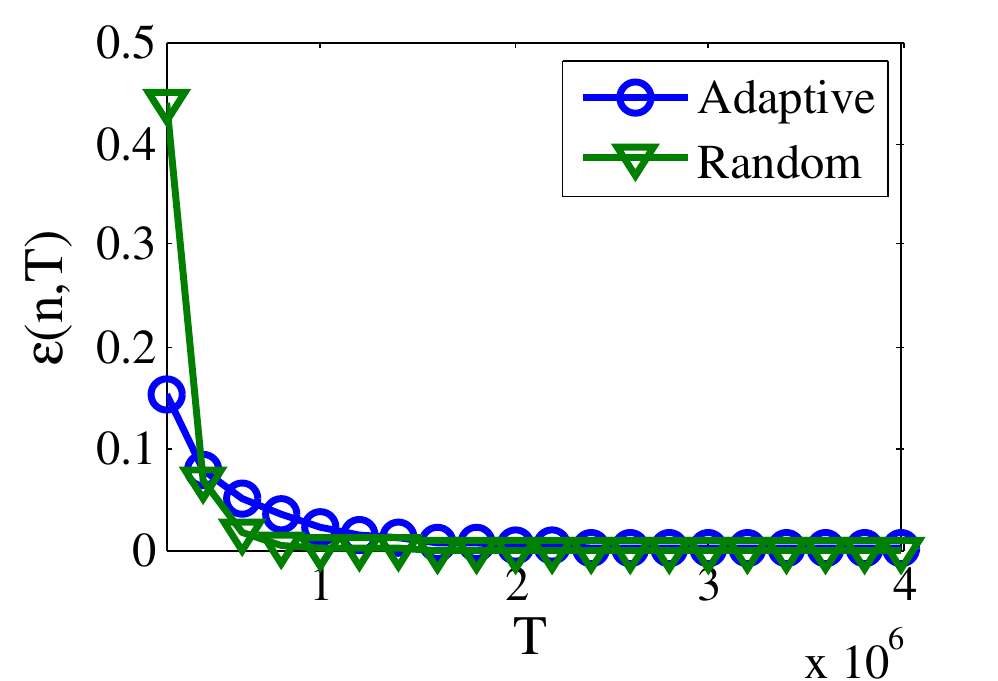} \label{fig:p3}
}
    \caption{The average fraction of misclassified nodes under SP (URS-1 sampling strategy) and ASP.}
    \label{fig:numerical}
  \end{center}
\end{figure*}

In this appendix, using toy examples, we numerically compare the performance of SP and ASP. We consider a network of 4000 nodes and two communities ($K=2$) with equal sizes ($\alpha=0.5$). In Figure \ref{fig:numerical}, we plot the average fraction of misclassified nodes as a function of the observation budget $T$. We use the URS-1 sampling strategy (refer to as `Random' in the experiments). As expected, the adaptive sampling algorithm outperforms non-adaptive algorithms, and in general, the performance increases with $\frac{(p-q)^2}{p}\frac{T}{n}$. 

Under URS-2 sampling strategy, we know, from \cite{heimlicher2012community}, that the fraction of misclassified nodes cannot be less than $1/2$ when $\frac{(p-q)^2}{p+q}\frac{T}{n} < 1.$ For example, if $p=10^{-3}$ and $q=p/20$, $T = 2.326\times 10^{6}$ is the required budget to be able to design algorithms that perform better than simply assigning nodes to clusters randomly. This is consistent with the results of our numerical experiments: for non-adaptive sampling, the fraction of misclassified nodes is less than 0.5 if $T\ge 2.6\times 10^6$ (roughly). In this case, adaptive sampling provides much better performance: the fraction of misclassified nodes is less than 0.5 if $T\ge 4\times 10^5$. This good performance achieved even with a small budget can be explained by the fact that we use a fraction $T/5$ of the budget on a small fraction of nodes ($n/5 \log n$ nodes) to identify kernels. The phase transition then occurs when $\frac{(p-q)^2}{p+q}\log n \frac{T}{n} < 1$, i.e., when $T=3\times 10^5.$


\section{Lower Bounds}

We derive the lower bounds on $\varepsilon(n,T)$ using a change-of-measure argument similar to those used in the bandit optimization literature \cite{lai1985} (i.e., we assume that the random observations are generated by a network whose structure is slightly different than the true structure). 

\subsection{Proof of Theorem~\ref{thm:lower-bound}}
  
Denote by $\Phi$ the true hidden partition $(V_k)_{1\le k \le K}$. Let
$\mathbb{P}_\Phi$ be the probability measure capturing the randomness
in the observations assuming that the network structure is described
by $\Phi$. We also introduce a slightly different structure
$\Psi$. The latter is described by clusters $V_1'=V_1 \setminus \{ v_1
\}$, $V_2'=V_2 \setminus \{ v_2 \}$, $V_k' = V_k$ for all $3 \le k \le
K$ and an isolated set $V_{12}'=\{ v_1 , v_2 \}$ with arbitrary
selected $v_1\in V_1$ and $v_2 \in V_2$. The observations intra- and
inter-cluster for nodes in $(V_k')_{1\le k \le K}$ are generated as in
the initial SBML, and for $v\in V_{12}'$ and for all $w \in V_1'\cup
V_2'$, when the node pair $(v,w)$ is observed, label $\ell$ is
observed with probability $\nu(\ell)$. For $v\in V_{12}'$ and for all
$w \notin V_1'\cup V_2'$, when the node pair $(v,w)$ is observed, label
$\ell$ is observed with probability $q(\ell)$.

Let $\pi\in \Pi'$ denote a clustering algorithm with output $(\hat{V}_k)_{1\le k \le K}$, and let $\set{E} = \bigcup_{1\le k \le K}\hat{V}_k \setminus V_k$ be the set of misclassified nodes under $\pi$. Note that in general in our proofs, we always assume without loss of generality that $| \bigcup_{1\le k \le K}\hat{V}_k \setminus V_k | \le |\bigcup_{1\le k \le K} \hat{V}_k \setminus V_{\sigma(k)}|$ for any permutation $\sigma$ of $\{1,\ldots,K\}$, so that the set of misclassified nodes is really $\set{E}$. Further define  $\set{B}=\{ v_1 \in \hat{V}_1, ~v_2 \in
\hat{V}_2\}$ as the set of events where nodes $v_1$ and $v_2$ are correctly classified. We can of course assume that $|\set{E}| \le \frac{(K-1)n}{K}$, and  we have $\varepsilon (n) = |\set{E}|$.  

Let $x_{i,j}$ denote the label observed on node pair $(i,j)$. We introduce $L$ (a quantity that resembles the log-likelihood ratio between $\mathbb{P}_\Phi$ and $\mathbb{P}_{\Psi}$) as: 

\begin{equation}L = \sum_{i \in V_1'}\log \frac{\nu(x_{i,v_1})\nu(x_{i,v_2})}{p(x_{i,v_1})q(x_{i,v_2})} + 
\sum_{i \in V_2'}\log \frac{\nu(x_{i,v_1})\nu(x_{i,v_2})}{q(x_{i,v_1})p(x_{i,v_2})},\end{equation}
In what follows, we establish a relationship between $\Ex [\varepsilon (n) ]$ and $L$. For any function $f(n)$, 
\begin{eqnarray} \label{eq:8-1}
\mathbb{P}_{\Psi} \{ L \le  f(n) \}& =& \mathbb{P}_{\Psi} \{ L \le  f(n) , \bar{\set{B}} \} + \mathbb{P}_{\Psi} \{ L \le  f(n) , \set{B} \}. 
\end{eqnarray}
We have: 
\begin{eqnarray}\mathbb{P}_{\Psi} \{ L \le  f(n) , \bar{\set{B}}\} &= &
  \int_{\{ L \le  f(n) , \bar{\set{B}}\}} d\mathbb{P}_{\Psi} \cr
&= & \int_{\{ L \le  f(n) , \bar{\set{B}}\}} \prod_{i \in
  V_1'}\frac{\nu(x_{i, v_1})\nu(x_{i,v_2})}{p(x_{i,v_1})q(x_{i,v_2})}
\prod_{i \in V_2'}\frac{\nu(x_{i,v_1})\nu(x_{i,v_2})}{q(x_{i, v_1})p(x_{i,v_2})} d\mathbb{P}_{\Phi} \cr
&\le & \exp (f(n))\mathbb{P}_{\Phi} \{ L \le  f(n) , \bar{\set{B}}\}~ \le ~ \exp (f(n))\mathbb{P}_{\Phi} \{ \bar{\set{B}}\}\cr
&\le &\frac{1}{\alpha_1}\exp (f(n))\Ex_{\Phi} [ \varepsilon (n)],
\label{eq:bdl-1} \end{eqnarray}
where the penultimate inequality comes from the fact that, since nodes in the
same community play identical roles,
\begin{eqnarray*}\mathbb{P}_{\Phi} \{ \set{B} \} &\ge& 1- \mathbb{P}_{\Phi} \{ v_1 \notin
\hat{V}_1 \} - \mathbb{P}_{\Phi} \{ v_2\notin
\hat{V}_2 \} \ge 1-\frac{1}{\alpha_1} \Ex_{\Phi}[ \varepsilon(n) ]. \end{eqnarray*}
We also have: 
\begin{equation}\mathbb{P}_{\Psi} \{ L \le  f(n) , \set{B} \} \le
  \mathbb{P}_{\Psi} \{  \set{B} \} = \frac{ \mathbb{P}_\Psi \{ v_1 \in \hat{V}_1, v_2
\in \hat{V}_2 \} +\mathbb{P}_\Psi \{ v_1 \in \hat{V}_2, v_2
\in \hat{V}_1 \}  }{2} \le
\frac{1}{2}.
\label{eq:bdlf-1}
\end{equation} 
Indeed $v_1$ and $v_2$ play identical roles under $\Psi$. Hence
$\mathbb{P}_\Psi \{v_1 \in \hat{V}_1, v_2 \in \hat{V}_2 \}$ and
$\mathbb{P}_\Psi \{ v_1 \in \hat{V}_2, v_2 \in \hat{V}_1 \}$ are
equal.
Combining \eqref{eq:8-1},~\eqref{eq:bdl-1}, and \eqref{eq:bdlf-1}, we get
$$\mathbb{P}_{\Psi} \{ L \le  f(n) \} \le \frac{1}{\alpha_1}\Ex_{\Phi}[\varepsilon(n) ]\exp(f(n)) +
\frac{1}{2}. $$

Since  $\Ex_{\Phi}[\varepsilon(n) ] =\Ex [\varepsilon(n) ],$ choosing $f(n) = \log
\left( \frac{\alpha_1}{4 \Ex [ \varepsilon(n) ]} \right)$, we obtain:\begin{equation}\label{eq:infimum}
 \lim\inf_{n \rightarrow \infty}\mathbb{P}_{\Psi} \{ L \ge  \log \left( \frac{\alpha_1}{4 \Ex [ \varepsilon(n) ]} \right) \} \ge \frac{1}{4}. 
\end{equation}

Now introduce
\begin{eqnarray*}L' & = &\sum_{i\in V_1'}
  \frac{\nu(x_{i, v_1 }) -
    p(x_{i,v_1})}{p(x_{i,v_1})}
  +\sum_{i\in V_1'}\frac{\nu(x_{i,v_2})-q(x_{i,v_2})}{q(x_{i,v_2})}
  \cr & &+ \sum_{i \in V_2'}
  \frac{\nu(x_{i, v_1})-q(x_{i,v_1})}{q(x_{i,v_1})}+\sum_{i\in V_2'}  \frac{\nu(x_{i,v_2})-p(x_{i,v_2})}{p(x_{i,v_2})}.\end{eqnarray*}
Then, $L \le L',$ since $\log (x) \le x-1.$ Thus,
\begin{equation}\label{eq:lll2}
 \lim\inf_{n \rightarrow \infty}\mathbb{P}_{\Psi} \{ L' \ge  \log \left( \frac{\alpha_1}{4 \Ex [ \varepsilon(n) ]} \right) \} \ge \frac{1}{4}. 
\end{equation}
By Chebyshev's inequality, we have 
$
\mathbb{P}_{\Psi}[L'\ge \mathbb{E}[L'] + 2\sigma_{\Psi}(L')]\le {1\over 4},
$
where $\sigma_{\Psi}[L']^2$ is the variance of $L'$ under $\Psi$. Hence, from (\ref{eq:lll2}), we deduce that: 
\begin{equation}\Ex_{\Psi}[L'] + 2\sigma_{\Psi}[L'] \ge \log \left(
\frac{\alpha_1}{4 \Ex [ \varepsilon(n) ]} \right).
\label{eq:chevy-1}
\end{equation}
 
We use the assumption of the sparse regime to deduce that $\lim_{n \to \infty} \sum_{\ell \neq 0} \frac{|p(\ell)-q(\ell)|}{\min\{p(0), q(0) \}} = 0$. Now we apply the above analysis with $\nu(\ell) = \min\{ p(\ell) , q(\ell)\}$ for all $\ell \neq 0.$ 
We have:
\begin{eqnarray}
\Ex_{\Psi}[L'] &\le&(\alpha_1 + \alpha_2) n\sum_{\ell}\frac{ \nu(\ell)}{p(\ell)}\big( \nu(\ell)-p(\ell) \big)+ (\alpha_1 + \alpha_2)n\sum_{\ell}\frac{ \nu(\ell)}{q(\ell)}\big( \nu(\ell)-q(\ell) \big)  \cr
& = &-(\alpha_1 + \alpha_2) n\sum_{\ell \neq 0}
\frac{\min\{p(\ell),q(\ell) \}}{\max\{p(\ell),q(\ell) \}}
|p(\ell)-q(\ell))| + (\alpha_1 + \alpha_2) n\frac{\nu(0)}{p(0)}(\nu(0)-p(0)) \cr&&+(\alpha_1 + \alpha_2)n\frac{\nu(0)}{q(0)}(\nu(0)-q(0)) \cr
& \le &2(\alpha_1 + \alpha_2)n\sum_{\ell }\frac{(p(\ell)-q(\ell))^2}{p(\ell) + q(\ell)} 
\label{eq:ek} 
\end{eqnarray}
Due to the fact that the $x_{i,j}$'s are independent, we also have: 
\begin{eqnarray}
\sigma_{\Psi}[L']^2& = &  \sum_{i\in V_1'}
  \sigma_{\Psi}\left[\frac{\nu(x_{i,v_1}) -
    p(x_{i,v_1})}{p(x_{i,v_1})}\right]^2
  +\sum_{i\in V_1'}\sigma_{\Psi}\left[\frac{\nu(x_{i,v_2})-q(x_{i,v_2})}{q(x_{i,v_2})} \right]^2
  \cr & &+ \sum_{i \in V_2'}
 \sigma_{\Psi}\left[
   \frac{\nu(x_{i,v_1})-q(x_{i,v_1})}{q(x_{i,v_1})}
   \right]^2+\sum_{i \in V_2'}  \sigma_{\Psi} \left[
    \frac{\nu(x_{i,v_2})-p(x_{i,v_2})}{p(x_{i,v_2})} \right]^2 \cr
&\le& (\alpha_1 + \alpha_2)n\sum_{\ell \neq 0}\frac{\nu(\ell)}{p(\ell)}\frac{\big( p(\ell)-\nu(\ell)
  \big)^2}{p(\ell)}+ (\alpha_1 + \alpha_2)n\sum_{\ell\neq 0}\frac{\nu(\ell)}{q(\ell)}\frac{\big( q(\ell)-\nu(\ell)
  \big)^2}{q(\ell)} \cr && + (\alpha_1 + \alpha_2)n\cdot\nu(0)\left(\frac{\nu(0)-p(0)}{
    p(0)}\right)^2 +(\alpha_1 + \alpha_2)n\cdot\nu(0)\left(\frac{\nu(0)-q(0)}{
    q(0)}\right)^2 \cr
&\le & 4(\alpha_1 + \alpha_2) n\sum_{\ell}\frac{(p(\ell)-q(\ell))^2}{p(\ell) +
  q(\ell)}, \label{eq:vak}
\end{eqnarray}
where the last inequality comes from the sparse regime assumption and from
$\frac{\nu(\ell)}{p(\ell)} \le 1.$ By \eqref{eq:chevy-1},
\eqref{eq:ek}, and \eqref{eq:vak},
$$2(\alpha_1 + \alpha_2 )\tau (n) + 4 \sqrt{(\alpha_1 + \alpha_2 ) \tau (n)} \ge \Ex_{\Psi}[L'] +2\sigma_{\Psi}[L']  \ge \frac{1}{2}\log \left( \frac{\alpha_1}{4 \Ex [
    \varepsilon(n) ]} \right) .$$
If $\lim_{n\rightarrow \infty} \Ex [ \varepsilon(n)] = 0,$ $\lim_{n\to
\infty}\tau(n) = \infty$. Thus, the above inequality becomes:
$$2(\alpha_1 + \alpha_2 ) \tau (n) \ge \frac{1}{2}\log \left( \frac{\alpha_1}{4 \Ex [
    \varepsilon(n) ]} \right). $$ 
This concludes the proof.

\subsection{Proof of Theorem~\ref{thm:l1}}

Without loss of generality, we assume that $KL(p,q) \ge
KL(q,p)$. The case where $KL(p,q) < KL(q,p)$ can be treated using exactly the same arguments, and is omitted. Alternatively, one may treat the case where $KL(p,q) < KL(q,p)$ by switching the roles played by $p$ and $q$, as well as the roles of observations with outcome 1 and those with outcome 0. More precisely, by such a change, $p$ is replaced by $1-q$, and $q$ by $1-p$. This simple argument explains why in the derived conditions and bounds, $p$ and $1-q$ play symmetric roles.

We follow exactly the same arguments of those used in the proof of Theorem~\ref{thm:lower-bound}. Under structure $\Psi$, for $v \in V_{12}'$ and for all $w$, when the node pair $(v, w)$ is observed, the outcome is equal to 1 with probability $q$. The label on a node pair here represents all the observations made on this pair. For example, the probability to observe a pair of nodes of the same cluster 4 times with 2 positive observations is $\eta(4)p^2(1-p)^2$, where $\eta(4)$ is the probability that the pair is observed 4 times. Refer to Section 3 in the paper for a detailed explanation of the correspondance. 

Now as in the previous proof, we obtain:
\begin{equation}
\Ex_{\Psi}[L] + 2\sigma_{\Psi}[L] \ge \log \left(\frac{\alpha_1}{4 \Ex [ \varepsilon(n,T) ]} \right).\label{eq:chevy}
\end{equation}

Let $m$ be a random variable with the same law as that of the number
of observations on pairs $(v , v_1)$  and
  $(w,v_2) $ over all $v\in V_1'$ and $w \in V_2'$. For both URS-1 and URS-2 sampling strategies, 
$$\Ex_{\Psi}[m] = T \frac{2((\alpha_1 + \alpha_2)n-2)}{n(n-1)} \quad \mbox{and}\quad
\sigma_{\Psi}[m]^2  \le T \frac{2((\alpha_1 + \alpha_2)n-2)}{n(n-1)}.$$
This is due to the facts that under URS-1, $m$ has a binomial distribution and that the variance of $m$ under URS-2 is smaller than the variance of $m$ under URS-1. Thus, 
\begin{equation}\label{eq:me2}
\Ex_{\Psi}[L] = \Ex_{\Psi}[m] KL(q,p) \le  \frac{2(\alpha_1 + \alpha_2)T}{n}KL(q,p). \end{equation}
From Lemma~\ref{lem:vark} (see below), we also deduce that:
\begin{align}\label{eq:sig2}
\sigma_{\Psi}[L]^2 \le  \frac{4(\alpha_1+\alpha_2)T}{n}\left(q\left(
    \log\frac{p(1-q)}{q(1-p)}\right)^2 + \left(
    \log\frac{1-q}{1-p}\right)^2 \right).
\end{align}
The first statement of the theorem is then obtained by combining (\ref{eq:chevy}), (\ref{eq:me2}), and (\ref{eq:sig2}).
\begin{multline}
\log \left(\frac{\alpha_1}{4 \Ex [ \varepsilon(n,T) ]} \right) \le \cr
 \frac{2(\alpha_1 + \alpha_2)T}{n}KL(q,p) + 2\sqrt{ \frac{4(\alpha_1+\alpha_2)T}{n}\left(q\left(
    \log\frac{p(1-q)}{q(1-p)}\right)^2 + \left(
    \log\frac{1-q}{1-p}\right)^2 \right)}. \label{eq:flbndna} \end{multline}

To prove the second statement of the theorem, we consider three cases: {\em a)} $p \le
Cq,$ {\em b)} $p >Cq$ and $1-p=\Theta(1)$, {\em c)} $1-p = o(1)$ with
constant $C$ such that $\frac{1}{C} \log^2(C) = \frac{1}{12}.$

\smallskip
\noindent\underline{\em  a) When $p \le Cq$:} Since we assume that $p \ge q$ and $KL(p,q)\ge KL(q,p)$, the from Lemma~\ref{lem:klpqqp}, when $p \le Cq,$ $1-p\ge q \ge \frac{p}{C}$. We observe that:
\begin{align}
\sigma_{\Psi}[L]^2 &\le  \frac{4(\alpha_1 + \alpha_2)T}{n}\left(q\left(
    \log\frac{p(1-q)}{q(1-p)}\right)^2 + \left(
    \log\frac{1-q}{1-p}\right)^2 \right) \cr
& \le  \frac{4(\alpha_1 + \alpha_2)T}{n}\left(    \frac{(p-q)^2}{q(1-p)^2} +
  \frac{(p-q)^2}{(1-p)^2} \right) \le \frac{8(\alpha_1 + \alpha_2)T}{n} \frac{(p-q)^2}{q(1-p)^2} \cr
& \le  \frac{8(\alpha_1 + \alpha_2)T}{n} (C+1)^3 \frac{(p-q)^2}{p+q}  \le \frac{16(\alpha_1 + \alpha_2)T}{n} (C+1)^3 KL(q,p),
\end{align}
where the last inequality results from Lemma~\ref{lem:klbounds} (see below). Therefore, when $\Ex [\varepsilon(n,T)]\rightarrow 0$, \eqref{eq:chevy} becomes
$$
 \frac{2(\alpha_1 + \alpha_2)T}{n}KL(q,p) \ge  \log \left(
  \frac{\alpha_1}{4 \Ex [ \varepsilon(n,T) ]} \right).
$$

\smallskip
\noindent\underline{\em \underline b)  When $p >Cq$ and
  $1-p=\Theta(1)$:} From Lemma~\ref{lem:klbounds}, $KL(q,p) \ge
\frac{p}{3}.$ From this, we can deduce
\begin{align}q \left(\log\frac{p(1-q)}{q(1-p)}\right)^2 &\le q \left(2\log\frac{p}{q}\right)^2 \le \frac{4}{C} \log^2(C)
p \le  KL(q,p) \quad \mbox{and} \cr \left( \log\frac{1-q}{1-p}\right)^2 ~~ &\le \frac{(p-q)^2}{(1-p)^2}
\le \frac{3}{(1-p)^2}KL(q,p) = \Theta(KL(p,q) ).\end{align} 
Therefore, when $\Ex[\varepsilon(n,T)]\rightarrow 0$, \eqref{eq:flbndna} becomes
$$ \frac{2(\alpha_1 + \alpha_2)T}{n}KL(q,p) \ge  \log \left(
  \frac{\alpha_1}{4 \Ex [ \varepsilon(n,T) ]} \right).$$

\smallskip
\noindent\underline{\em c) When $1-p = o(1)$:} By
Lemma~\ref{lem:klpqqp}, $q \le 1-p$. Therefore, $KL(q,p) = \Theta
(\log\frac{1-q}{1-p}).$ Thus, for this case, \eqref{eq:flbndna} becomes
$$\frac{-\log \mathbb{E}[\varepsilon (n,T)]}{ (\frac{T}{n} +
  \sqrt{\frac{T}{n}} ) KL(q,p)} = O(1). $$
When $\frac{-\log \mathbb{E}[\varepsilon (n,T)]}{ KL(q,p)} = \omega(1),$
$\frac{T}{n} \to \infty$ and \eqref{eq:flbndna} becomes
$$ \frac{2(\alpha_1 + \alpha_2 )T}{n}KL(q,p) \ge  \log \left(
  \frac{\alpha_1}{4 \Ex [ \varepsilon(n,T) ]} \right).$$
When $\frac{-\log \mathbb{E}[\varepsilon (n)]}{ KL(q,p)} = O(1),$
In this case, the condition from
\eqref{eq:flbndna} is $\frac{T}{n} = O(1).$ But this bound is not
tight. When $\Ex[\varepsilon(n,T)]\rightarrow 0$, the observed edges should generate a giant component containing almost every nodes. With
random sampling, the condition to get the giant component is $\frac{T}{n} = \omega(1).$

From the three cases, we can conclude that when
$\mathbb{E}[\varepsilon (n,T)] = o(1)$, $\frac{T}{n} = \omega(1)$ and
$$ \frac{2(\alpha_1 + \alpha_2)T}{n} \ge  \frac{-\log \left(\frac{4}{\alpha_1} \Ex [
    \varepsilon(n,T) ] \right)}{KL(q,p)}.$$

\begin{lemma} $\min \{ KL(q,p), KL(p,q)\} \ge \frac{(p-q)^2}{2(p+q)}$.
\label{lem:klbounds}
\end{lemma}

{\it Proof.} From the definition of $KL(q,p),$
\begin{align*}
KL(q,p)  &=  q \log \frac{q}{p} + (1-q) \log \frac{1-q}{1-p} = -q \log \frac{p}{q} - (1-q) \log \frac{1-p}{1-q} \cr
  &= -q \log\left(1+ \frac{p-q}{2q} \right)-q \log\left(1+ \frac{p-q}{p+q} \right) - (1-q) \log \frac{1-p}{1-q} \cr
 &\ge -q  \frac{p-q}{2q}-q \frac{p-q}{p+q}  - (1-q) \frac{q-p}{1-q} =(p-q)\left(\frac{1}{2} - \frac{q}{p+q} \right)  =  \frac{(p-q)^2}{2(p+q)}.
\end{align*}

\begin{lemma} $KL(p,q) \ge KL(q,p)$ iff $p(1-p) \ge
  q(1-q).$ \label{lem:klpqqp} \end{lemma}

{\it Proof.} 
Let $p = \frac{1}{2}+cx$ and $q = \frac{1}{2}+c,$ where $c\in
[-{1\over 2},\frac{1}{2}].$ We will show that $KL(p,q)-KL(q,p)
\ge 0$ when $-1\le x\le 1$. Let,
$$g(x)=KL(p,q)-KL(q,p) = (1+(1+x)c)\log\frac{1+2cx}{1+2c} + (1-(1+x)c)\log\frac{1-2cx}{1-2c}. $$
Then, $g'(1)=0,$ $g(-1)=g(1)=0$, and
\begin{eqnarray*}
g'(x) &= & c\log\frac{1+2cx}{1+2c}+\frac{2c(1+(1+x)c)}{1+2cx} -c\log\frac{1-2cx}{1-2c}-\frac{2c(1-(1+x)c)}{1-2cx} \cr
g''(x) &= & 4c^3 (x-1) \left(  \frac{1}{(1+2cx)^2} - \frac{1}{(1-2cx)^2} \right). 
\end{eqnarray*}
From this, we will deduce  $g(0)\ge 0$.  First, one can easily check $g(0)=0$ when $c=0.$ Let $h(c) =  -(1+c)\log(1+2c)
- (1-c)\log(1-2c).$ Then, $h'(c) =  \log \frac{1-2c}{1+2c}
-\frac{2+2c}{1+2c} + \frac{2-2c}{1-2c} $ and
 $h''(c) = 4c\left( \frac{1}{(1-2c)^2}-\frac{1}{(1+2c)^2} \right) \ge
 0.$ Since $h'(0) = 0$ and $h''(c) \ge 0$, $h(c) \ge 0$. Therefore,
 $g(0) \ge 0.$

We now conclude the proof of the lemma. When $-1 \le x \le 0,$ since $g(-1)=0$, $g(0)\ge 0$, and $g''(x) \le 0$,  $g(x) \ge 0.$ For $0\le x \le 1,$ since $g(0)\ge 0$, $g(1)=0$, $g'(1)=0$ and $g''(x) \le 0$, $g(x) \ge 0.$
Therefore, $KL(p,q) \ge KL(q,p)$ iff $p(1-p) \ge q(1-q).$

\begin{lemma}When $KL(p,q)\ge KL(q,p)$, then there exists constant $C$ such that if $p \ge Cq$, 
$$KL(p,q) \ge \frac{p}{3}\log\left(\frac{p(1-q)}{q(1-p)} \right) .$$\label{lem:klubnd}
\end{lemma}

{\it Proof.} From Lemma~\ref{lem:klpqqp}, $\frac{p}{q} \ge
\frac{1-q}{1-p}.$ Thus, when $p\ge \frac{5}{6}$
$$KL(p,q) = p\log\left(\frac{p(1-q)}{q(1-p)} \right) -
\log\left(\frac{1-q}{1-p} \right) \ge (p-\frac{1}{2})  \log\left(\frac{p(1-q)}{q(1-p)}
\right) \ge \frac{p}{3}\log\left(\frac{p(1-q)}{q(1-p)} \right). $$

When $p \le \frac{5}{6}$ and $C$ is sufficiently large,
$\log(\frac{1-q}{1-p})$ becomes much smaller than $
p\log\left(\frac{p}{q} \right) .$ Thus,
$$KL(p,q) = p\log\left(\frac{p}{q} \right) +(1-p)
\log\left(\frac{1-p}{1-q} \right) \ge \frac{2p}{3}\log\left(\frac{p}{q} \right)  \ge \frac{p}{3}\log\left(\frac{p(1-q)}{q(1-p)} \right). $$

\begin{lemma} 
$\sigma_{\Psi} [L]^2 \le 2(\sigma_{\Psi} [m])^2 \left( \log\frac{1-q}{1-p}\right)^2+2\big( \Ex_{\Psi} [m] q(1-q) +q^2(\sigma_{\Psi} [m])^2 \big) \left( \log\frac{p(1-q)}{q(1-p)}\right)^2.$ 
 \label{lem:vark}
\end{lemma}

{\it Proof.} Let $s$ denote a random variable representing the number of positive observations gathered on one the following pairs: $(1,\alpha n), \dots,(\alpha n-1,\alpha n),$ and  $(\alpha n+1,n)\dots (n-1,n)$. Then,
\begin{align*}
\Ex_{\Psi} \left[ \left( L- \Ex_{\Psi} [L]\right)^2 \right]  & = \Ex_{\Psi} \Big[ \Big( (m - \Ex_{\Psi} [m]) \log\frac{1-q}{1-p} -(s-\Ex_{\Psi} [s]) \log \frac{p(1-q)}{q(1-p)} \Big)^2 \Big] \cr
& \le  2\Ex_{\Psi} \left[ (m - \Ex_{\Psi} [m])^2
  \left(\log\frac{1-q}{1-p}\right)^2 \right] +2\Ex_{\Psi} \left[(s-\Ex_{\Psi} [s])^2 \left(\log \frac{p(1-q)}{q(1-p)}\right)^2
\right] \cr
& =  2 (\sigma_{\Psi} [m])^2
  \left(\log\frac{1-q}{1-p}\right)^2 
+2(\sigma_{\Psi} [s])^2 \left(\log \frac{p(1-q)}{q(1-p)}\right)^2.
\end{align*}

Now if $\eta(m,s)$ denote the joint probability distribution of $m$ and
$s$, we have:
\begin{eqnarray*}
(\sigma_{\Psi} [s])^2  &=&  \sum_{m \ge 1}
\sum_{s=0}^{m}(s^2-\Ex_{\Psi} [s]^2) \eta(m,s) \cr 
&= & \sum_{m \ge 1} \sum_{s=0}^{m}(s^2-m^2 q^2 +m^2 q^2 - \Ex_{\Psi} [s]^2) \eta(m,s) \cr
&= & \sum_{m \ge 1} \sum_{s=0}^{m}(s^2-m^2 q^2) \eta(m,s) + \sum_{m \ge 1} \sum_{s=0}^{m}(m^2 q^2 - \Ex_{\Psi} [s]^2) \eta(m,s) \cr
& = & \sum_{m \ge 1} mq(1-q)\sum_{s=0}^{m} \eta(m,s) + \sum_{m \ge 1} \sum_{s=0}^{m}(m^2 q^2 - \Ex_{\Psi} [m]^2q^2) \eta(m,s) \cr 
&= & q(1-q) \Ex_{\Psi} [m]  +q^2 (\sigma_{\Psi}[m])^2. 
\end{eqnarray*}
The lemma easily follows.

\subsection{Proof of Theorem~\ref{thm:la}}
As in the previous proof, without loss of generality, we assume that $KL(p,q) \ge KL(q,p)$.

Let $d(v)$ denote the number of samplings on node $v$. Again we use a
change-of-measure argument to show the following lemma, which gives
the lower bound of $d(v)+d(w)$ when we arbitrary select $v$ and $w$
from different clusters. The proof of the lemma will be given after
this proof.
\begin{lemma} For any $v_i \in V_i$, $v_j \in V_j$ and $i\neq j$, $\mathbb{P}_{\Psi} \left\{d(v_i)+d(v_j) \ge  \frac{1}{2p}
\right\} \ge \frac{1}{2}$. In addition,
when $\frac{- p \log \mathbb{E}[\varepsilon (n,T)]}{KL(p,q) }  =
\omega( 1 )$, $\lim\inf_{n
\rightarrow \infty } \mathbb{P}_{\Psi} \left\{ 
d(v_i)+d(v_j) \ge \frac{-\log \mathbb{E}[\varepsilon (n,T)]}{4KL(p,q)}\right\}=1$.\label{lem:pdd}
\end{lemma}

To conclude this proof, we have to show \eqref{eq:univ3} and
\eqref{eq:univ4} for $p$ and $q$ following $\frac{-\log
  \mathbb{E}[\varepsilon^{\pi}(n,T)]}{\log\frac{p}{q}} = \omega(1)$
and \eqref{eq:univ3} for $p$ and $q$ following $\frac{-\log
  \mathbb{E}[\varepsilon^{\pi}(n,T)]}{\log\frac{p}{q}} = O(1)$.

\smallskip
\noindent{\em a) Consider $\frac{-\log
  \mathbb{E}[\varepsilon^{\pi}(n,T)]}{\log\frac{p}{q}} = \omega(1)$.}
Since
 $KL(p,q)  = O( p\log \frac{p}{q} )$,
$\frac{-\log \mathbb{E}[\varepsilon^\pi(n,T)]}{ \log \frac{p}{q} } =
\omega(1)$ implies that $\frac{-p \log \mathbb{E}[\varepsilon^\pi(n,T)]}{
  KL(p,q) } = \omega(1)$. Then, by Lemma~\ref{lem:pdd},
\begin{eqnarray}
T &= &\sum_{k=1}^K\sum_{v\in V_k}
\frac{d(v)}{2} 
~\ge~ \min\{\frac{1}{2},1-\alpha_n \}\frac{n}{8 KL(p,q)} \log \frac{1}{ \Ex [\varepsilon (n,T) ]}, \label{eq:udbnd} \end{eqnarray}
since nodes in the same cluster play the same role and there are $\min\{\frac{1}{2},1-\alpha_n \}n$ non-overlapping pairs of nodes from different clusters.  
We have proved (\ref{eq:univ4}). Now we establish
(\ref{eq:univ3}). Note that the second condition in (\ref{eq:univ3})
is a direct consequence of (\ref{eq:univ4}). Further observe that by
assumption, $\frac{-\log
  \mathbb{E}[\varepsilon^{\pi}(n,T)]}{KL(p,q)} = \omega(1)$. Together with (\ref{eq:udbnd}), this implies that ${pT\over n}=\omega(1)$, and hence the first condition in (\ref{eq:univ3}) holds.  

\smallskip
\noindent{\em b) Consider $\frac{-\log
  \mathbb{E}[\varepsilon^{\pi}(n,T)]}{\log\frac{p}{q}} = O(1)$.} In
this case, $\log\frac{p}{q} = \omega(1)$ which means
$p=\omega(q)$. By combining this and Lemma~\ref{lem:klubnd},
$KL(p,q)=\Theta (p\log\frac{p}{q})$. Thus, to prove \eqref{eq:univ3}, it
is enouhg to show that
$p\frac{T}{n} = \Omega(1)$, since $\frac{T}{n}KL(p,q) =
\Theta(\frac{T}{n}p\log\frac{p}{q})$ and $\log \frac{p}{q} = \omega(1)$.  By Lemma~\ref{lem:pdd},
\begin{eqnarray}
T &= &\sum_{k=1}^K\sum_{v\in V_k}
\frac{d(v)}{2} 
~\ge~ \min\{\frac{1}{2},1-\alpha_n \}\frac{n}{8 p}. \label{eq:udbnd2} \end{eqnarray}
From the above ineqaulity, ${T\over n} = \Omega( \frac{1}{p}).$

\subsection{Proof of Lemma~\ref{lem:pdd}}
In this proof, we provide the proof for $i=1$ and $j=2$. Other cases
can be shown analogously.

Let $\Phi$ be the true
network structure: $(V_k )_{1\le k \le K}$. The modified structure $\Psi$ changes two
clusters: ${V}'_1 = \{v_2 \} \cup V_1 \setminus \{v_1 \}$, ${V}'_2
= \{v_1 \} \cup V_2 \setminus \{ v_2 \}$, and $V_k' = V_k$ for $3\le k
\le K$, where $v_1 \in V_1$ and $v_2 \in V_2$. The difference between $\Phi$ and $\Psi$
concerns only the nodes $v_1$ and $v_2$. 

We denote by $e(t)$ the edge selected in round $t$ by the adaptive sampling algorithm. Let $\ell(t)\in\{0,1\}$ be the outcome of the observation in round $t$. Define $p(1)=p$, $q(1)=q$, $p(0)=1-p$, and $q(0)=1-q$. Further define: for $\ell\in \{0,1\}$, for any node pair $e$:
$$
\mu (\ell, e, \Phi) = \begin{cases} p(\ell) & \mbox{if}~e \in \set{S}_{\Phi}
  \cr q(\ell) & \mbox{if}~e \notin \set{S}_{\Phi}   \end{cases}$$
and
$$ \mu (\ell, e, \Psi) = \begin{cases} p(\ell) & \mbox{if}~e \in \set{S}_{\Psi}
  \cr q(\ell) & \mbox{if}~e \notin \set{S}_{\Psi} \end{cases}$$
where $\set{S}_{\Phi}$ and $\set{S}_{\Psi}$ denote the sets of node pairs such that both nodes are in the same cluster under $\Phi$ and $\Psi,$ respectively.

Define
\begin{align}
L &= \sum_{t=1}^{T}
  \log\frac{\mu(\ell(t),e(t),\Psi)}{\mu(\ell(t),e(t),\Phi)} \cr
&= \sum_{t=1}^{T} (1_{v_1 \in e(t)} + 1_{v_2 \in e(t)})\log\frac{\mu(\ell(t),e(t),\Psi)}{\mu(\ell(t),e(t),\Phi)} ,
\end{align}
where $1_{a \in e}$ is equal to 1 when node pair $e$ contains node $a$, and to 0 otherwise. We further define $d(a) = \sum_{t=1}^{T}1_{a \in e(t)}.$

Since nodes in the same community
play identical roles, $\mathbb{P}_{\Phi} \{ v_1 \notin \hat{V}_1 \}
\le \frac{1}{\alpha_1}\Ex [ \varepsilon (n,T) ]$ and $\mathbb{P}_{\Psi} \{ v_1 \notin \hat{V}_2 \} \le \frac{1}{\alpha_2} \Ex [ \varepsilon (n,T) ]$.
Define the event $C=\{  v_1 \notin \hat{V}_1, d(v_1)+d(v_2) \le \zeta,  L\le  f(n)\}$. Then:
\begin{eqnarray}
\frac{1}{\alpha_1} \Ex_{\Phi}[\varepsilon (n,T) ]~\ge~\mathbb{P}_{\Phi}\left\{v_1 \notin \hat{V}_1\right\}  &\ge& \mathbb{P}_{\Phi}  \{ C \} ~ =~  \int_{C}
\prod_{t = 1
  }^{T}\frac{\mu(\ell(t),e(t),\Phi)}{\mu(\ell(t),e(t),\Psi)}
d\mathbb{P}_{\Psi} \cr
 &\ge& \exp(-f(n)) \mathbb{P}_{\Psi} \{ C \}.\label{eq:err}
\end{eqnarray}
Choosing $f(n) = -\frac{1}{2}\log  \Ex [\varepsilon (n,T) ],$
from \eqref{eq:err}, we obtain: $\lim\sup_{n\rightarrow \infty} \mathbb{P}_{\Psi} \left\{C\right\} =0$.
Therefore, since $\lim\sup_{n\rightarrow \infty} \mathbb{P}_{\Psi}
\left\{C\right\} =0$ and  $\lim\inf_{n\rightarrow \infty}\mathbb{P}_{\Psi} \{v_1 \notin \hat{V}_1\} \ge 1-\lim\sup_{n\to\infty}\frac{\Ex [\varepsilon(n,T)]}{\alpha_2} = 1,$ 
$$
\lim_{n \rightarrow \infty }\mathbb{P}_{\Psi} \left\{ 
d(v_1)+d(v_2) \le \zeta  , L
\le  f(n)\right\} = 0,
$$
which means that
\begin{align}
\lim_{n \rightarrow \infty }\mathbb{P}_{\Psi} \left\{ 
d(v_1)+d(v_2) \le \zeta , L \ge  f(n)\right\}  + \lim_{n \rightarrow \infty }\mathbb{P}_{\Psi} \left\{ 
d(v_1)+d(v_2) > \zeta \right\} = 1 .\label{eq:dbnd1}
\end{align}

Let $x (j)$ and $e(j)$ denote the value and the edge of the $j$-th
observation such that $v_1 \in e(j)$ or $v_2 \in e(j)$. We define $L(t)$ as the $L$ but computed up to $t$-th observation :
\begin{align*}L(t) = & \sum_{j=1}^t 1_{e(j)  \in \mathcal{S}_{\Psi}} \cdot \left( x(j) \log(\frac{p(1-q)}{q(1-p)}) +
\log(\frac{1-p}{1-q}) \right)+\cr &\sum_{j=1}^t 1_{e(j)  \notin \mathcal{S}_{\Psi}} \cdot \left(x(j)
\log(\frac{q(1-p)}{p(1-q)}) +
\log(\frac{1-q}{1-p})\right).\end{align*} 
Then,
\begin{equation}\mathbb{P}_{\Psi} \left\{ d(v_1)+d(v_2) \le  \zeta   , L\ge  f(n) \right\}  \le
\mathbb{P}_{\Psi} \left\{ \sup_{t \le  \zeta } L (t)
  \ge f(n) \right\}.\label{eq:supl}\end{equation}

To complete the proof of the first part of this lemma, we set $\zeta =\frac{1}{2p}$. When
$\sum_{j =1 }^{t} x(j) = 0$ and $t \le \frac{1}{2p},$ 
$$L(t) \le t \log \frac{1-q}{1-p} \le \frac{1}{2p} \log \frac{1-q}{1-p} \le
\frac{1}{2p}\frac{p-q}{1-p} < f(n).$$
Therefore, by \eqref{eq:supl}
\begin{eqnarray}
\mathbb{P}_{\Psi} \left\{ d(v_1)+d(v_2) \le  \frac{1}{2p}   , L\ge  f(n) \right\} &\le&
\mathbb{P}_{\Psi} \left\{ \sup_{d\le \frac{1}{2p}} L(d) \ge  f(n)
\right\}\cr & \le &1- P_{\Psi} \left\{ \sum_{j =1 }^{1/2p} x(j)  = 0  \right\} \cr &\le& 1-(1-p)^{1/2p}\le
\frac{1}{2}.\label{eq:lbndc1}
\end{eqnarray}
By combining \eqref{eq:dbnd1} and \eqref{eq:lbndc1},
$$\mathbb{P}_{\Psi} \left\{ d(v_1)+d(v_2) \ge  \frac{1}{2p} \right\}
\ge \frac{1}{2}.$$

In what follows, to conclude this proof, we will consider  when $\frac{- p \log \mathbb{E}[\varepsilon (n,T)]}{KL(p,q) }  =
\omega( 1 )$ and show that $\lim_{n \rightarrow \infty }\mathbb{P}_{\Psi} \left\{ 
\sup_{t \le \frac{f(n)}{2KL(p,q)}} L(t) \ge  f(n)\right\} = 0$.
Then, from \eqref{eq:dbnd1} and \eqref{eq:supl}, we can deduce that
$$\lim\inf_{n\to \infty}\mathbb{P}_{\Psi} \left\{ d(v_1)+d(v_2) \ge \frac{f(n)}{2KL(p,q)}  \right\}=1.$$
 For this case, we set $\zeta = \frac{f(n)}{2KL(p,q)}.$
Then $\zeta = \omega(1).$ 

To bound $\mathbb{P}_{\Psi} \left\{ \sup_{t \le \zeta} L(t) \ge  f(n)\right\}$, we use Doob's maximal inequality.
\begin{lemma}[Doob's maximal inequality] Let $\{ L(t) \}$ be a
  martingale such that $ (\sigma [L(T)])^2 < \infty$. Then,
$$\mathbb{P}\{ \max_{1\le t \le T} |L(t) | \ge \lambda \}
\le \frac{1}{\lambda^2}  (\sigma [L(T)])^2 .$$\label{lem:kolmo}
\end{lemma}

We slightly modify $L (t)$ so that we can use
Lemma~\ref{lem:kolmo}. Let  
\begin{align*}L_m(t) = & \frac{1}{\zeta KL(p,q)}\sum_{j=1}^t 1_{e(j)  \in \mathcal{S}_{\Psi}} \cdot \left( x(j) \log(\frac{p(1-q)}{q(1-p)}) +
\log(\frac{1-p}{1-q})-KL(p,q) \right)+\cr 
&\frac{1}{\zeta KL(p,q)}\sum_{j=1}^t 1_{e(j)  \notin \mathcal{S}_{\Psi}} \cdot \left(x(j)
\log(\frac{q(1-p)}{p(1-q)}) +
\log(\frac{1-q}{1-p})-KL(q,p)\right).\end{align*} 
On may easily check that $L_m(t)$ is a martingale, and for $t \le \zeta$,
\begin{equation}L(t) - L_m(t) \zeta KL(p,q) \le \zeta KL(p,q) =
  \frac{f(n)}{2}. \label{eq:llmd}\end{equation}
The variance of increments of $L_m(t)$ is bounded as follows:
\begin{align*}(\sigma_{\Psi}[L_m(t+1)-L_m (t)])^2 &\le \frac{\max\{p(1-p),q(1-p)\}}{\zeta^2KL(p,q)^2}\log^2(\frac{p(1-q)}{q(1-p)})\cr
&=\frac{p(1-p)}{\zeta^2KL(p,q)^2}\log^2(\frac{p(1-q)}{q(1-p)}),\end{align*}
since $p(1-p) \ge
q(1-p)$ when $KL(p,q) \ge KL(q,p)$ by
Lemma~\ref{lem:klpqqp}. Therefore, the variance of $L_m(\zeta)$ satisfies
\begin{equation}(\sigma_{\Psi}[L_m (\zeta) ])^2 \le \frac{ p(1-p)
    \log^2(\frac{p(1-q)}{q(1-p)})}{\zeta KL(p,q)^2}
  . \label{eq:sigmaL} \end{equation} 
Let $C$ be a large enough constant such that $KL(p,q) \ge
\frac{p}{3}\log(\frac{p(1-q)}{q(1-p)})$ when $p >Cq$. The existence of such constant $C$ is ensured in view of Lemma~\ref{lem:klubnd}.  Now,
consider two cases: $p \le Cq$ and $p > Cq$.

When $p \le Cq$, $1-p \le q \le \frac{p}{C}$  by
Lemma~\ref{lem:klpqqp} and $p \le \frac{C}{C+1}$. For this case,
\eqref{eq:sigmaL} becomes
\begin{eqnarray*} 
(\sigma_{\Psi}[L_m (\zeta) ])^2 & \le& \frac{ p(1-p)
    \log^2(\frac{p(1-q)}{q(1-p)})}{\zeta KL(p,q)^2} \cr
&\le &\frac{p \frac{(p-q)^2}{q^2(1-p)} }{\zeta KL(p,q)^2} ~\le~\frac{C(C+1)\frac{(p-q)^2}{q} }{\zeta KL(p,q)^2} ~\le~  \frac{4C(C+1)^2}{\zeta  KL(p,q)},
\end{eqnarray*}
where we used Lemma \ref{lem:klbounds} for the last
inequality. Therefore, with $\zeta = \frac{f(n)}{2 KL(p,q)},$
since $f(n)$ tends to infinity, $ (\sigma_{\Psi}[L_m
(\zeta) ])^2 = o(1)$.

When $p > Cq$, since $KL(p,q) \ge \frac{p}{3}\log\left(\frac{p(1-q)}{q(1-p)}\right),$
$(\sigma_{\Psi}[L_m (\zeta) ])^2 \le
\frac{3(1-p)}{\zeta p} .$
Since in case {\em a)}, we assume $\frac{ pf(n)}{KL(p,q) }  =
\omega( 1 )$, then with $\zeta = \frac{f(n)}{2 KL(p,q)},$
$(\sigma_{\Psi}[L_m (\zeta) ])^2 = o(1).$

Next we apply Lemma~\ref{lem:kolmo}. Since $(\sigma_{\Psi}[L_m (\zeta) ])^2 = o(1)$ when $\frac{ pf(n)}{KL(p,q) }  =
\omega( 1 )$, by \eqref{eq:llmd} and Lemma~\ref{lem:kolmo},
$$\lim_{n \to \infty}\mathbb{P}_{\Psi} \left\{ \sup_{t \le
   \zeta} L(t)  \ge f(n) \right\} \le \lim_{n \to \infty}\mathbb{P}_{\Psi} \left\{ \sup_{t \le  \zeta} L_m (t)  \ge 1 \right\} =0 .$$

\section{Spectral Partition}
\subsection{Proof of Theorem~\ref{thm:specdecom}}
In what follows, we use the standard matrix norm $\| A\|=\sup_{\|
  x\|=1}\| Ax\|$. $X_{\Gamma} = A_{\Gamma} - \mathbb{E}[A_{\Gamma}]$ is a random matrix
of which elements have zero mean. Let $\delta = \frac{\alpha_1}{10}$. This proof proceeds as following:
we 
bound $\| X_{\Gamma} \|$ from random matrix theory,  show that
 the fraction of misclassified nodes on $\Gamma$ goes to 0 with
the bounded $\| X_{\Gamma} \|$, and conclude this proof by proving that with high probability
$|V\setminus \Gamma| = o(n).$ 

We first show that, with high probability,  $\| X_{\Gamma} \| =
O(\sqrt{p\frac{T}{n}})$.  
\begin{lemma}
With high probability, the following condition holds:
\begin{itemize}
\item[(C3)] $\| X_{\Gamma} \| \le \sqrt{C_1 (p+q)\frac{T}{n}}$ for
  some constant $C_1$, where
  $X_{\Gamma}=A_{\Gamma}-\mathbb{E}[A_{\Gamma}]$.
\end{itemize}
\label{lem:FOspectral}
\end{lemma}
The proof of the above lemma is postponed to the end of this section. The proof of Lemma \ref{lem:FOspectral} relies on arguments used in the spectral analysis of random graphs \cite{feige2005spectral}.

Since the eigen values of
$\mathbb{E}[A_{\Gamma}]$ are the order of $\Omega((p-q)\frac{T}{n})$,
when $\frac{(p-q)^2}{p}\frac{T}{n}= \omega(1)$, $\|X_{\Gamma} \|$ is
negligible to the eigen values of $\mathbb{E}[A_{\Gamma}]$. Thus, by
spectral decomposition, we can recover $\mathbb{E}[A_{\Gamma}]$ from
given $A_\Gamma$ and reconstruct the comunities. In following lemmas, we bounds $|\bigcup_{1\le k \le K}(S_k \setminus
V_k)\cap \Gamma|$ under (C3) :  
\begin{lemma} When $|V \setminus \Gamma | < \delta n ,$ under (C3), after Algorithm~\ref{alg:trimspec}, we have:
$$
|(S_1 \bigtriangleup V_1)\cap \Gamma|  \le   \frac{1-\alpha}{\alpha}\left(\frac{4\sqrt{C_1 (p+q)\frac{T}{n}}}{(p-q)\alpha\frac{T}{n} - \sqrt{C_1 (p+q)\frac{T}{n}}}\right)^2 n.
$$
\label{lem:spect}
\end{lemma}
\begin{lemma} 
When $|V \setminus \Gamma | < \delta n ,$ under (C3), after
Algorithm~\ref{alg:specg}, we have: with sufficiently large constant $C_3$,
$$
|\bigcup_{1\le k \le K}(S_k \setminus V_k)\cap \Gamma|  \le   C_3 \frac{p}{(p-q)^2}\frac{n^2}{T}.
$$ \label{lem:spectgk}
\end{lemma}

Therefore, if we show that $|V\setminus \Gamma| = o(n)$ with high probability, since (C3)
also occurs with high probability, Lemma~\ref{lem:spect} and
Lemma~\ref{lem:spectgk} imply that, when $\frac{(p-
  q)^2}{p}\frac{T}{n} = \omega(1)$,
$$\lim_{n \to \infty} \frac{1}{n}|\bigcup_{1\le k \le K}(V_k \setminus
S_k)| = \lim_{n \to \infty} \frac{1}{n}|\bigcup_{1\le k \le K}(S_k \setminus
V_k)\cap \Gamma| +\frac{1}{n}|V \setminus \Gamma| = 0.$$

Now, we will complete this proof by showing that $|V\setminus \Gamma|
= o(n)$ with high probability. By law of large numbers, $5K
\frac{\sum_{(v,w) \in E } A_{vw}}{n} \ge 4 p\frac{T}{n}$. When  $\frac{(p-
  q)^2}{p}\frac{T}{n} = \omega(1)$, since $\mathbb{E}[\sum_{w\in V}A_{vw}] \le 2p\frac{T}{n}$
and $(\sigma[\sum_{w\in V} A_{vw}])^2 \le 2p\frac{T}{n}$, by Chebychev's
inequality, we can show that all $v\in V$  satisfy that
$\lim_{n\to \infty}\mathbb{P}\{ v\in \Gamma\} = 1$. From this, it is
also true that $|V\setminus \Gamma|
= o(n)$ with high probability.

\subsection{Proof of Lemma~\ref{lem:FOspectral}}
Let $X_{V_{i}\times V_{j} ,\Gamma}$ denote the matrix constructed from $X_{\Gamma}$ by keeping the entries $(i,j)$ in $(V_{i}\cap \Gamma) \times (V_{j} \cap \Gamma)$. To establish the lemma, we prove that 
with high probability,
\begin{equation}\label{eq:lll}
\| X_{V_{i}\times V_{j} ,\Gamma} \| \le
\frac{1}{K^2}\sqrt{C_1(p+q)\frac{T}{n}} \quad \mbox{for all} \quad 1\le
i,j \le K.
\end{equation}
Then from the convexity of matrix norm,
\begin{eqnarray}
\| X_{\Gamma}\|&\le& \sum_{1\le i,j \le K} \| X_{V_{i}\times V_{j} ,\Gamma}\|~\le~\sqrt{C_1(p+q)\frac{T}{n}}.
\end{eqnarray}
Next we prove (\ref{eq:lll}) for $i=j=1$. The proof consists in extending arguments in \cite{feige2005spectral}. We provide
the proof for $X_{V_{1}\times V_{1} ,\Gamma}$ under URS-1 sampling strategy. Other cases can be shown analogously. We first introduce the notion of {\it discrepancy}, an extension of a similar notion used in \cite{feige2005spectral}. Let $e(A,B)$  denote the number of positive observations between nodes $(v,w)$ with $v\in A$ and $w\in B$ and let $\mu(A,B)$ denote the average of $e(A,B)$. Let $|V_1 \cap \Gamma| = n'$. 
\begin{definition}[Discrepancy property]
All $A, B \subset V_1\cap\Gamma$ satisfy one of the following properties: for some constants $c_2$ and $c_3$,
\begin{itemize}
\item$e(A,B)/\mu (A,B) \le
 c_2$ 
\item$e(A,B) \log(e(A,B)/\mu (A,B)) \le c_3 \max{(|A|,|B|)} \log
 (n'/\max{|A|,|B|}).$
\end{itemize}
\end{definition}

As in \cite{feige2005spectral}, it can be shown, using arguments from random matrix theory, that if the discrepancy property holds, then (\ref{eq:lll}) holds. Next we establish that with high probability, the discrepancy property holds. Let $A,B$ be two subsets of $V_1\cap\Gamma$, and $c_2,c_3$ two large enough constants. Without loss of generality we assume that $| B|\ge |A|$. We distinguish two cases:

\smallskip
\noindent(Case1: if $|B| \ge n'/5$.) Due to the trimming step in SP, for all $v\in V_1\cap\Gamma,$ $e(v,V_1)
\le 10Kp\frac{T}{n}.$ Thus. 
$$e(A,B) \le |A|\cdot 10Kp\frac{T}{n} \le c_2  \mu(A,B).$$ 

\smallskip
\noindent(Case2: if $|B| \le n'/5$) 
Let $\eta(A,B) = \max\{\eta^{\star}, c_2 \mu(A,B) \}$ where
$\eta^{\star}$ is the constant satisfying that $ \eta^\star \log(\eta^\star/\mu (A,B)) - c_3 |B| \log
(n'/|B|) =0 .$ Then, if all pairs $A,B \subset V_1 \cap \Gamma$ satisfy
$e(A,B) \le \eta(A,B)$, the discrepancy property holds. Thus, we just show that with
high probability $e(A,B) \le \eta(A,B)$ for all $A,B \subset V_1 \cap \Gamma$.

First, we quantify the probability that $e(A,B) \le \eta(A,B)$
for any arbitrary subsets $A$ and $B$ of $V_1$. Between $A$ and $B$, there are at most $|A||B|$ and at least
$\frac{|A|(|B|-1)}{2}$ node pairs.  Under URS-1, on $i$-th sampling, each
node pair is selected with probability $\frac{2}{n(n-1)}$ and the
positive observation appears with probability $p.$ Thus,
$\frac{|A|(|B|-1)}{n(n-1)}pT \le \mu (A,B) \le \frac{2 |A||B|}{n(n-1)}pT .$
Then, by Markov inequality,
\begin{eqnarray}
\mathbb{P}\{ e(A,B) > \eta(A,B) \} & \le & \inf_{h \ge 0}\frac{\mathbb{E}[\exp (h \cdot  e(A,B))]}{\exp(\eta(A,B) h)} \cr
& \le & \inf_{h \ge 0} \frac{\prod_{t=1}^T (1+\frac{2p |A||B|}{n(n-1)}  \exp (h))}{\exp(\eta(A,B) h)}  \cr
& \le & \inf_{h \ge 0} \frac{\prod_{t=1}^T \exp(   \frac{2 p |A||B|
  }{n(n-1)} \exp(h))}{\exp(\eta(A,B) h)} \cr
& \le & \inf_{h \ge 0} \exp(  4\exp(h) \mu (A,B)  -\eta(A,B) h  ) \cr
& \le & \exp( -\eta (\log \frac{\eta(A,B)}{4 \mu (A,B)}-1)),\label{eq:c2pre}
\end{eqnarray}
where, for the last inequality, we set $h=\log \frac{\eta(A,B)}{4 \mu (A,B)}.$

Then, we compute the expected number of pairs $A,B \subset V_1
\cap \Gamma$ such that $e(A,B) > \eta(A,B)$. The number of possible pairs of sets $A$ and $B$ such that $A\in V_1$ and $B \in V_1$ with size $|A|=a$ and
$|B|=b$ is ${{n'}\choose{a}} {{n'}\choose{b}}$. Hence using \eqref{eq:c2pre}, 
\begin{align}
\mathbb{E}[|\{(A,B): & e(A,B)> \eta(A,B),~ |A|=a,~|B|=b,~A,B\subset
V_1\cap \Gamma  \}|]
\cr &  \le  {{n'}\choose{a}} {{n'}\choose{b}}\exp\left( -\eta(A,B) (\log
  \frac{\eta(A,B)}{4 \mu (A,B)}-1)\right) \cr
& \stackrel{(a)}{\le}   \left(\frac{n'e}{b}\right)^{2b} \exp\left( -\eta(A,B) (\log
  \frac{\eta(A,B)}{4 \mu (A,B)}-1)\right)\cr
&  \stackrel{(b)}{\le}  \exp\left(4b\log\frac{n'}{b} -\eta(A,B) (\log
  \frac{\eta(A,B)}{4 \mu (A,B)}-1)\right)\cr
&\le \exp\left(-3\log n + 3\log n + 4b\log\frac{n'}{b} -\eta(A,B) (\log
  \frac{\eta(A,B)}{4 \mu (A,B)}-1)\right) \cr
&\stackrel{(c)}{\le} \exp\left(-3\log n + 7b\log\frac{n'}{b} -\eta(A,B)
  (\log \frac{\eta(A,B)}{4 \mu (A,B)}-1)\right)\cr
&\stackrel{(d)}{\le} \exp\left(-3\log n + 7b\log\frac{n'}{b}
  -\frac{\eta(A,B)}{2} (\log \frac{\eta(A,B)}{\mu
    (A,B)})\right) \stackrel{(e)}{\le}\frac{1}{n^3}.,\label{eq:etabnd}
\end{align}
where for $(a)$ and $(b)$, we use $b \le \frac{n'}{5}$; for $(c)$, we use that $x \log x$ is an increasing function; $(d)$ stems from $\frac{\eta(A,B)}{\mu (A,B)} \ge c_2$; and $(e)$ is obtained by the definition of $\eta(A,B)$. Therefore, by summing the above inequality for all possible cardinalities $a,b$, we get: 
$\mathbb{E}[|\{(A,B) : e(A,B)> \eta(A,B),~A,B\subset
V_1\cap \Gamma \}|] \le \frac{1}{n}$ and we can conclude that with high probability the discrepancy property holds.

\subsection{Proof of Lemma~\ref{lem:spect}}
 For notational simplicity, let $m= | \Gamma |$ and $ |V_1 \cap \Gamma | =
  \tilde{\alpha} m.$  Let $\tilde{x}_1$ and $\tilde{x}_2$
  be two orthonormal  vectors defined by $\tilde{x}_1 (v) =
  \frac{1}{\sqrt{m}}$ for all $v\in \Gamma$ and $\tilde{x}_2 (v)=
 \sqrt{ \frac{1-\tilde{\alpha}}{m\tilde{\alpha}}}$ for $ v \in V_1 \cap \Gamma$ and $ \tilde{x}_2 (v)=
  -\sqrt{\frac{\tilde{\alpha}}{(1-\tilde{\alpha})m}}$ for $v \in V_2\cap
  \Gamma$.  We define
  $\tilde{x}_i $ for $3 \le i \le m $ so that $\{\tilde{x}_i \}_{1\le
    i \le m}$ is an orthonormal basis of $\mathbb{R}^m .$

  We denote by $\lambda_i$ and $x_i$, $i$-th largest eigenvalue and
  corresponding eigenvector of $A_{\Gamma}$ and denote by
  $\tilde{\lambda}_1$ and $\tilde{\lambda}_2$ the two eigenvalues of $\mathbb{E}[A_{\Gamma}].$ 
The eigenvalues of $\mathbb{E}[A_{\Gamma}]$ are 
\begin{align*}\tilde{\lambda}_1 &= \left( p+\sqrt{p^2 - 4\tilde{\alpha}(1-
     \tilde{ \alpha} ) (p^2 - q^2)}\right) \frac{mT}{n^2} ,\cr
\tilde{\lambda}_2 &= \left(p-\sqrt{p^2 - 4\tilde{\alpha} (1-\tilde{\alpha}) (p^2 -
    q^2)}\right) \frac{mT}{n^2}.\end{align*}
Observe that $\tilde{\lambda}_2 \ge 2(p-q)\tilde{\alpha} \frac{mT}{n^2}$. This is due to the fact that if$f(\tilde{\alpha}) = \left((1-2\tilde{\alpha}) p + 2\tilde{\alpha} q \right)^2 -\left( p^2 -4\tilde{\alpha}(1-\tilde{\alpha})(p^2 - q^2)\right)$, then $f(\tilde{\alpha})\le 0$ (indeed, $f(0)=f(1/2)=0$ and $f''(x) = -16q(p-q)>0,$ $f(\tilde{\alpha})\le 0$ for $0 \le \tilde{\alpha} \le \frac{1}{2}.$) $x_1$ and $x_2$ can be represented as $x_1 = \sum_{i=1}^{m} \gamma_i \tilde{x}_i \quad \mbox{and} \quad x_2 = \sum_{i=1}^{m} \theta_i \tilde{x}_i,$ where $\sum_{i=1}^{m} \gamma_i^2 =1$ and $\sum_{i=1}^{m} \theta_i^2 =1$.

From the definition of $x_1,$
\begin{equation}
  A_{\Gamma} x_1  =  \lambda_1 x_1 = \lambda_1\sum_{i=1}^{m}
  \gamma_i \tilde{x}_i ,\label{eq:ax1}
\end{equation}
and from the definition of the othornormal basis $\{\tilde{x}_i \},$
\begin{equation}
 A_{\Gamma}x_1  = (\mathbb{E}[A_{\Gamma}]+X_{\Gamma})x_1=
 X_{\Gamma}x_1 + \mathbb{E}[A_{\Gamma}] \sum_{i=1}^{2}\gamma_i \tilde{x}_i. \label{eq:ax2}
\end{equation}
Since $\| X_{\Gamma} x_1\| \le \| X_{\Gamma} \|,$ when we combine \eqref{eq:ax1}
and \eqref{eq:ax2},
\begin{eqnarray}
\|X_{\Gamma}\|^2 & \ge & \| \lambda_1\sum_{i=1}^{m} \gamma_i \tilde{x}_i  - M_{\Gamma} \cdot \sum_{i=1}^{2}\gamma_i \tilde{x}_i  \|^2 
 ~ \ge~  \| \lambda_1\sum_{i=3}^{m} \gamma_i \tilde{x}_i \|^2 ~=~
 \lambda_1^2 \sum_{i=3}^{m} \gamma_i^2 \cr
& \ge & (\tilde{\lambda}_1 - \|X_{\Gamma}\|)^2\sum_{i=3}^{m}
\gamma_i^2 , \label{eq:xgamma}
\end{eqnarray}
where for the last inequality, we use: $\lambda_i \ge \tilde{\lambda}_i - \|X_{\Gamma}\|$.
Similarly, we can show that
\begin{equation}\|X_{\Gamma}\|^2 \ge  (\tilde{\lambda}_2 -
\|X_{\Gamma}\|)^2\sum_{i=3}^{m} \theta_i^2 .\label{eq:xgamma2}\end{equation}

By definition of $\hat{x}$ (see Algorithm 2), $\hat{x} = x_1 + x_2 - (\gamma_1 + \theta_1)
\tilde{x}_1.$ Define $z = \hat{x}-(\gamma_2 + \theta_2 )\tilde{x}_2 = \sum_{i=3}^{m} (\gamma_i + \theta_i) \tilde{x}$. We classify node $v$ by looking at the sign of $\hat{x}(v)$, whereas the true classification is determined by the sign of $\tilde{x}_2(v)$. Then since $|\tilde{x}_2 (v)| \ge \sqrt{\frac{\tilde{\alpha}}{(1-\tilde{\alpha})m}}$, $v$ is misclassified only if $|z(v)| \ge (\gamma_2 + \theta_2 ) \sqrt{\frac{\tilde{\alpha}}{(1-\tilde{\alpha})m}}$. Therefore, $\| z\|^2$ is greater than the product of the number of misclassified nodes and of $(\gamma_2 + \theta_2 ) \sqrt{\frac{\tilde{\alpha}}{(1-\tilde{\alpha})m}}$. In other words,
\begin{eqnarray*}
\frac{| (S_1 \bigtriangleup V_1)\cap \Gamma |}{|
  \Gamma |} \frac{\tilde{\alpha}(\gamma_2 + \theta_2 )^2}{1-\tilde{\alpha}} & \le & \|z\|^2 ~ = ~
\sum_{i=3}^m (\gamma_i + \theta_i)^2 ~ \le ~ 2\sum_{i=3}^m \gamma_i ^2+ 2\sum_{i=3}^m \theta_i^2\cr&  \le &
\frac{2\|X_{\Gamma}\|^2}{(\tilde{\lambda}_1 - \|X_{\Gamma}\|)^2} + \frac{2\|X_{\Gamma}\|^2}{(\tilde{\lambda}_2 - \|X_{\Gamma}\|)^2},
\end{eqnarray*}
where the last inequality uses \eqref{eq:xgamma} and \eqref{eq:xgamma2}.
Since $(\gamma_2 + \theta_2)^2 \ge 1-
\frac{2\|X_{\Gamma}\|^2}{(\tilde{\lambda}_1 - \|X_{\Gamma}\|)^2} -
\frac{2\|X_{\Gamma}\|^2}{(\tilde{\lambda}_2 - \|X_{\Gamma}\|)^2},$ the
above inequality becomes
\begin{eqnarray*}
\frac{| (S_1 \bigtriangleup V_1)\cap \Gamma |}{|
  \Gamma|}  & \le & \frac{1-\tilde{\alpha}}{\tilde{\alpha}}\left(\frac{4\|X_{\Gamma}\|^2}{(\tilde{\lambda}_1 - \|X_{\Gamma}\|)^2} + \frac{4\|X_{\Gamma}\|^2}{(\tilde{\lambda}_2 - \|X_{\Gamma}\|)^2}\right) \cr
& \le & \frac{1-\alpha_1}{\alpha_1} \frac{16\|X_{\Gamma}\|^2}{(\tilde{\lambda}_2 - \|X_{\Gamma}\|)^2} \cr
& \le & \frac{1-\alpha_1}{\alpha_1}\left( \frac{4\|X_{\Gamma}\|}{(p-q)\alpha_1\frac{T}{n} - \|X_{\Gamma}\|} \right)^2 .
\end{eqnarray*}

\subsection{Proof of Lemma~\ref{lem:spectgk}}
For notational simplicity, let $m= | \Gamma |$ and $ |V_k \cap \Gamma | =
  \tilde{\alpha}_k m.$
Let $M_{\Gamma}^{(k)}$ denote the column vector of $M_{\Gamma}$ on $v
\in V_{k} \cap \Gamma$ and $\gamma(k) = \arg\min_{k}\| \xi_{i^\star ,
  k}-M^{(k)} \|$. This proof will show that 
\begin{equation}\frac{1}{n}\left|\bigcup_{1\le k \le K}(S_k \setminus V_{\gamma(k)})\cap
  \Gamma\right| \le C_3 \frac{p}{(p-q)^2}\frac{n}{T} . \label{eq:svnorm}\end{equation}
Then, since $\frac{(p-q)^2}{p}\frac{T}{n} = \omega(1)$, the right hand
side of the above equation goes to 0. It is necessary that $\gamma(k)
= k$ so that $\frac{|S_k \setminus V_{\gamma(k)}|}{n} = 0$ for all
$k$. Therefore, \eqref{eq:svnorm} concludes this lemma.

\noindent {\bf Proof of \eqref{eq:svnorm}:} Let $M_{v,\Gamma}$ denote
the column vector of $M_\Gamma$ on $v$. Since $\|M_{v,\Gamma}
-M_{\Gamma}^{\gamma(k)} \|^2 \ge (p-q)^2
\frac{4T^2}{n^4}(\tilde{\alpha}_1 + \tilde{\alpha}_2)m$ for all $v \in
(S_k \setminus V_{\gamma(k)})\cap \Gamma$,
\begin{align}
\left|\bigcup_{1\le k \le K}(S_k \setminus V_{\gamma(k)})\cap \Gamma\right| &
(p-q)^2 \frac{4T^2}{n^4}(\tilde{\alpha}_1 + \tilde{\alpha}_2)m \cr &\le \sum_{k=1}^K
\sum_{v \in (S_k \setminus V_{\gamma(k)})\cap \Gamma} \|M_{v,\Gamma} -
M_{\Gamma}^{\gamma(k)} \|^2 \cr
&\le 2\sum_{k=1}^K\sum_{v \in (S_k \setminus V_{\gamma(k)})\cap \Gamma}
(\|M_{v,\Gamma} -\xi_{i^\star ,k}\|^2+ \|\xi_{i^\star ,k}- M_{\Gamma}^{\gamma(k)} \|^2 )\cr
&\le 4\sum_{k=1}^K\sum_{v \in (S_k \setminus V_{\gamma(k)})\cap \Gamma}
\|M_{v,\Gamma} -\xi_{i^\star ,k}\|^2 \cr
&\le 8\sum_{k=1}^K\sum_{v \in (S_k \setminus V_{\gamma(k)})\cap \Gamma}
(\|M_{v,\Gamma} - \hat{A}_{v}\|^2+\|\hat{A}_{v}- \xi_{i^\star ,k}\|^2 )\cr
&\le 8 \|M_{\Gamma}-\hat{A} \|_F^2 + 8 r_{i^\star},\label{eq:svbound}
\end{align}
where $\| \cdot \|_{F}$ denotes Frobenious norm. To complete the proof
of \eqref{eq:svnorm}, It is enough to
show the followings:
\begin{eqnarray}
8 \|M_{\Gamma}-\hat{A} \|_F^2 & \le & C_3 p (\tilde{\alpha}_1 + \tilde{\alpha}_2) \frac{T}{n} \qquad\mbox{and} \label{eq:frobbnd} \\
8r_{i^\star} & \le & C_3 p(\tilde{\alpha}_1 + \tilde{\alpha}_2)\frac{T}{n}, \label{eq:rstarbnd}
\end{eqnarray}

\noindent{\bf Proof of~\eqref{eq:frobbnd}:} This proof starts from the
property of Frobenius norm: when the rank of matrix $A$ is $K$,
$\|A\|_F^2 \le K \|A\|^2$. Since the ranks of $\hat{A}$ and $M_{\Gamma}$
are $K$, the rank of $\hat{A} - M_{\Gamma}$ is less than $2K$. Thus,
\begin{eqnarray}
\sum_{v\in \Gamma}\|\hat{A}_{v} - M_{v,\Gamma} \|^2 ~ = ~
\|\hat{A} - M_{\Gamma} \|_F^2 & \le & 2K \|\hat{A} -
M_{\Gamma} \|^2 \cr
&\le & 4 K \|\hat{A} - A_{\Gamma}\|^2 +
4K\|A_{\Gamma}-M_{\Gamma} \|^2 \cr
&\le& 8K \| A_{\Gamma} - M_{\Gamma}\|^2 \cr
&\le & 8K C_1 (p+q)\frac{T}{n}. \label{eq:frobam}
\end{eqnarray}
Therefore, when we let $C_3 \ge \frac{128KC_1}{\tilde{\alpha}_1 + \tilde{\alpha}_2},$
we can deduce \eqref{eq:frobbnd}.

\noindent{\bf Proof of~\eqref{eq:rstarbnd}:} It is sufficient to show
that there exists $i$ such that $r_i \le \frac{1}{8} C_3 p(\tilde{\alpha}_1 +
\tilde{\alpha}_2)\frac{T}{n} .$

Since $\frac{\sum_{(v,w) \in E } A_{vw}}{2n^2} \le p\frac{T}{n^2}$ by
law of large numbers, with constant $C_2>\frac{1}{\delta}$, there exists $i_t$ such that
$32K C_1 C_2(p+q)\frac{T}{n^2} \le i_t \frac{\sum_{(v,w) \in E } A_{vw}
}{100n^2} \le  64K C_1 C_2(p+q)\frac{T}{n^2}$ and $1\le i_t \le \log
n$. In what follows, we will show that $r_{i_t} \le \frac{1}{8} C_3 p(\tilde{\alpha}_1 +
\tilde{\alpha}_2)\frac{T}{n} .$

To complete this proof, we first bound
$\|\xi_{i_t,k}-M_{\Gamma}^{(k)}\|$. Let 
\begin{eqnarray*}
I_k & = &\{ v \in V_k \cap \Gamma : \|\hat{A}_v - M_{\Gamma}^{(k)} \|^2 \le \frac{1}{4}i_t \frac{\sum_{(v,w) \in E } A_{vw}
}{100n^2} \} \cr
O & = &\{ v \in \Gamma : \|\hat{A}_v - M_{\Gamma}^{(k)}
\|^2 \ge 4 i_t \frac{\sum_{(v,w) \in E } A_{vw}}{100n^2}~\mbox{for all}~ 1\le
k\le K \}.
\end{eqnarray*}
For all $v\in I_k,$ $I_k \subset Q_{i_t , v}$, since $\|\hat{A}_v - \hat{A}_w \|^2 \le
2\|\hat{A}_v -M_{\Gamma}^{(k)}\|^2 + 2\|\hat{A}_w
-M_{\Gamma}^{(k)}\|^2 \le i_t \frac{\sum_{(v,w) \in E } A_{vw}
}{100n^2} $ for all $w \in I_k$.  
Besides, for all $v\in O$, $|(\cup_{k=1}^{K} I_k) \cap
Q_{i_t , v} | =0$, since $\|\hat{A}_v - \hat{A}_w \|^2 \ge
\frac{1}{2}\|\hat{A}_v -M_{\Gamma}^{(k)}\|^2 - \|\hat{A}_w
-M_{\Gamma}^{(k)}\|^2 > i_t \frac{\sum_{(v,w) \in E } A_{vw}
}{100n^2} $ for all $w \in I_k$. Since from \eqref{eq:frobam}
\begin{eqnarray*}
|\Gamma \setminus (\cup_{k=1}^KI_k) | &\le&  8K C_1 (p+q)\frac{T}{n} \left(\frac{1}{4}i_t \frac{\sum_{(v,w) \in E } A_{vw}
}{100n^2}\right)^{-1}\cr
&\le&\frac{n}{C_2} ~\le ~ \delta n,
\end{eqnarray*}
$|Q_{i_t , v}| \le \delta n$ for all $v \in O$ and $|Q_{i_t , v}| \ge
\alpha_1 n - 2\delta n $ for all $v \in \cup_{k=1}^K I_k$.
Therefore, all $v \in O$ cannot be the origin of $T_{i_t, k}$. Since
the column vector of the origin of $T_{i_t , k}$ should be within
$\sqrt{4 i_t \frac{\sum_{(v,w) \in E } A_{vw}}{100n^2}}$ from
$M_{\Gamma}^{(k)}$ in Euclidean
distance,  we can deduce that
$$\| \xi_{i_t,k} - M_{\Gamma}^{(k)}\|^2 \le C_2 i_t \frac{\sum_{(v,w) \in E } A_{vw}
}{100n^2} \le 64KC_1(C_2)^2(p+q)\frac{T}{n^2}.$$
Therefore, for sufficiently large constant $C_3$, 
\begin{eqnarray*}
r_{i_t} & =& \sum_{1\le k \le K}\sum_{v \in T_{i_t,k}} \| \hat{A}_v
-\xi_{i_t,k} \|^2 \cr
&\le& \sum_{1\le k \le K}\sum_{v \in V_{k}\cap \Gamma} \| \hat{A}_v
-\xi_{i_t,k} \|^2 \cr
&\le& 2\sum_{1\le k \le K}\sum_{v \in V_{k} \cap \Gamma} ( \| \hat{A}_v
-M_{v,\Gamma}\|^2 +\|M_{v,\Gamma}-\xi_{i_t,k} \|^2 )\cr
&\le&16K C_1 (p+q)\frac{T}{n}+ K(C_1C_2)^2(p+q)\frac{T}{n} \cr
&\le&\frac{1}{8} C_3 p(\alpha_1 +\alpha_2)\frac{T}{n} .
\end{eqnarray*}

\subsection{Proof of Theorem~\ref{thm:algorithms}}

We first introduce the following notations: let
$e(v,S) = \sum_{w \in S} A_{vw}$ denote the total number of positive
observations for node pairs including node $v$ and a node from
$S$. Let $\delta=\alpha_1/10$. Further define $H$ as the (largest) set
of nodes $v$ satisfying:
\begin{itemize}
\item[(H1)] For all $k,$ $|e(v,V_k) -\Ex
  [e(v,V_k)] | \le
  \frac{p-q}{4} |V_k| \frac{2T}{n^2},$

\item[(H2)] $e(v,V\setminus H) \le
  \frac{p-q}{8}\frac{\alpha_1 T}{n}.$
\end{itemize}
We bound the size of $H$ in the follwoing lemma.
\begin{lemma}
When $\frac{(p-q)^2}{p}{\alpha_1 T\over n} =\omega(1)$ and
$\frac{(p-q)^2}{20p}{\alpha_1 T\over n} \ge \log(p{T\over n})$, with
high probability, 
$$|V \setminus H| \le \exp\left( -    \frac{(p-q)^2}{20p}\frac{\alpha_1 T}{n} \right)n .$$
\label{lem:sizeH}
\end{lemma}

In this proof, we will show that all nodes in $H$ are well classified
after SP algorithm. Then, by Lemma~\ref{lem:sizeH}, with high
probability, 
$$\varepsilon^{SP} (n,T) \le \exp\left( -  \frac{(p-q)^2}{20p}\frac{\alpha_1 T}{n} \right).$$

 Remember that after the trimming and the spectral
decomposition steps, the algorithm returns clusters $(S_k)_{1\le k
  \le K}$. According to Theorem~\ref{thm:specdecom}, with high probability, 
$$\frac{1}{n}|\bigcup_{k=1}^K V_k \setminus S_k| =
\frac{1}{n}|\bigcup_{k=1}^K (S_k \setminus V_k) \cap \Gamma| +
\frac{1}{n}|V\setminus \Gamma| =o(1). $$
Note that with high probability $H\subset \Gamma$ (this is a consequence of the law of large numbers in
view of the definitions of the two sets) and$\frac{1}{n}|V\setminus
H| = o(1)$ (from Lemma~\ref{lem:sizeH}).
Therefore, with high probability:
\begin{equation}\label{eq:step1}
|\bigcup_{1\le k \le K}(S_k\setminus V_k)\cap H | + |V \setminus
H | =o(n).
\end{equation}

 We now analyse the gains achieved in the ``improvement'' step. We start from (\ref{eq:step1}), and prove:

\begin{lemma}  When
${| \bigcup_{k=1}^K (S^{(0)}_k \setminus V_k)\cap H| + |V \setminus
  H|} \le \delta n,$
$$ \frac{ | \bigcup_{k=1}^K (S^{(i+1)}_k \setminus V_k)\cap H|}{
  |\bigcup_{k=1}^K (S^{(i)}_k \setminus V_k)\cap H|} \le \left( \frac{\|X_{\Gamma} \|}{ \frac{\alpha_1 T}{n}(p-q)\left(\frac{(\alpha_1 - \delta)^2}{\alpha_1(\alpha_1 + \delta)} - \frac{5}{8}  \right) } \right)^2.$$ \label{lem:improve}
\end{lemma}

From this result, we deduce that with high probability (using (C1) and $\frac{(p-
  q)^2}{p}\frac{T}{n} = \omega(1)$):
\begin{eqnarray*} 
\frac{ | \bigcup_{k=1}^K (S^{(i+1)}_k \setminus V_k)\cap H|}{
  |\bigcup_{k=1}^K (S^{(i)}_k \setminus V_k)\cap H|} &\le& \left( \frac{\|X_{\Gamma} \|}{ \frac{\alpha_1 T}{n}(p-q)\left(\frac{(\alpha_1 - \delta)^2}{\alpha_1(\alpha_1 + \delta)} - \frac{5}{8}  \right) } \right)^2 ~ \le~ e^{-2}.
\end{eqnarray*}
Therefore, after $\log n$ steps, $|\bigcup_{k=1}^K (S^{(\log n)}_k \setminus V_k)\cap H|
\le \frac{1}{n^{2}} n < 1$. We conclude that with high probability: $|\bigcup_{k=1}^K (\hat{V}_k \setminus V_k)\cap H |= 0.$ This means that with high probability, all nodes of $H$ are well classified. 

It remains to establish the various intermediate lemmas.

\subsection{Proof of Lemma~\ref{lem:sizeH}}
 We denote by $Z_1$ the set of nodes that do not satisfy (H1). 
 We show that $|Z_1|\le \frac{1}{2} \exp \left(-\frac{(p-q)^2 }{20p}\frac{\alpha_1 T}{n}\right)n$ with high probability. 

Observe that, when $v \in V_k$, $e(v,V_k)$ follows a binomial
  distribution Bin$(T, \frac{2 \alpha_k p}{n})$ and,  when $v \notin V_k$, $e(v,V_k)$ follows a binomial
  distribution Bin$(T, \frac{2 \alpha_k q}{n})$. Applying Chernoff's inequality  (as in Lemma 8.1 of \cite{coja2010}),
\begin{eqnarray*}
\mathbb{P}\left\{\frac{\left|e(v,V_k) -\Ex
      [e(v,V_k)] \right|}{|V_k|} \le
  \frac{p-q}{4}\frac{2T}{n^2} \right\}& \le& 2 \exp \left(
-\frac{\left(\frac{(p-q)\alpha_1 T}{2n}  \right)^2 }{2\left(
    \frac{2\alpha_1 p T }{n} +
    \frac{(p-q)\alpha_1 T}{6n} \right) } \right) \cr
&\le& 2 \exp \left(-\frac{(p-q)^2 }{18p } \frac{\alpha_1 T}{n}\right). 
\end{eqnarray*}
Therefore, 
$$\Ex [| Z_1 |] \le 2K \exp \left(-\frac{(p-q)^2 }{18p } \frac{\alpha_1
    T}{n}\right) n. $$
Applying Markov inequality,
$$\mathbb{P} \left\{ |Z_1 |\ge \frac{1}{2} \exp \left(-\frac{(p-q)^2 }{20p } \frac{\alpha_1
    T}{n}\right)n \right\} \le 4K\exp \left(-\frac{(p-q)^2 }{180p } \frac{\alpha_1
    T}{n}\right). $$

Hence from the assumptions made in Theorem \ref{thm:algorithms}, $| Z_1|\le  \frac{1}{2} \exp \left(-\frac{(p-q)^2 }{20p } \frac{\alpha_1 T}{n}\right)n$ with high probability. 
    
Next we prove the following intermediate claim: there is no subset
$S\subset V$ such that $e(S,S) \ge \frac{p-q}{16}\frac{\alpha_1
  T}{n}|S|$ and $|S|= \exp \left(-\frac{(p-q)^2 }{20p } \frac{\alpha_1
    T}{n}\right) n$ with high probability. For any subset $S \in V$
such that $|S| =  \exp \left(-\frac{(p-q)^2 }{20p } \frac{\alpha_1
    T}{n}\right) n,$ by Markov inequality,
\begin{eqnarray} 
\mathbb{P} \{ e(S,S) \ge \frac{p-q}{16}\frac{\alpha_1 T}{n}|S| \} &\le&\inf_{t\ge  0} \frac{ \mathbb{E}[ \exp (e(S,S)t) ]  }{ \exp \left( \frac{p-q}{16}
  \frac{\alpha_1 T}{n} |S|t \right) } \cr
&\le&  \inf_{t\ge 0} \frac{ \prod_{i=1}^T ( 1+\frac{|S|^2}{n^2} p \exp(t) )  }{ \exp\left( \frac{p-q}{16}  \frac{\alpha_1 T}{n} |S| t \right) } \cr
&\le & \inf_{t\ge 0} \exp \left( \frac{|S|^2}{n^2}pT \exp (t) -   \frac{p-q}{16}
  \frac{\alpha_1 T}{n} |S| t  \right) \cr
&\le &  \exp \left(   -   \frac{p-q}{16}
  \frac{\alpha_1 T}{n} |S|\left( \log \left(\frac{\alpha_1 (p-q) n}{16  p |S|}-1
  \right)  \right) \right)\cr 
&\le& \exp \left(   -   \frac{p-q}{16}
  \frac{\alpha_1 T}{n} |S| \right),\label{eq:bndss}
\end{eqnarray}
where, in the last two inequalities, we have set $t= \log \left( \frac{\alpha_1 (p-q) n}{16 p |S|} \right)$ and use the fact that: 
$$\log \left(  \frac{\alpha_1 (p-q) n}{16 p |S|} \right) = \log \left(
  \frac{\alpha_1 (p-q)^2 }{16 p }\frac{T}{n} \left((p-q)\frac{T}{n} \right)^{-1} \exp\left(
    \frac{(p-q)^2}{20p}\frac{\alpha_1 T}{n}\right) \right) >2,$$
which comes from the assumptions made in the theorem ($\frac{(p-q)^2}{20p}\frac{\alpha_1 T}{n} \ge \log(
(p-q)\frac{T}{n})$ and $\frac{(p-q)^2}{p}\frac{\alpha_1 T}{n} = \omega(1)$). 
Since the number of subsets $S \subset V$ with size $|S|$ is
${{n}\choose{|S|}} \le (\frac{e n}{|S|})^{|S|} ,$ from \eqref{eq:bndss}, we deduce:
\begin{align*} \mathbb{E}[|& \{S : e(S,S) \ge \frac{p-q}{16}\frac{\alpha_1 T}{n}|S|~\mbox{and}~|S|= \exp \left(-\frac{(p-q)^2 }{20p } \frac{\alpha_1
    T}{n}\right) n \} |] \cr 
&\le \exp\left(|S| + \frac{(p-q)^2 }{20p } \frac{\alpha_1
    T}{n}|S| \right) \exp\left( -\frac{p-q}{16} \frac{\alpha_1 T}{n} |S|
\right) \cr
&\le \exp\left(-\frac{(p-q)^2 }{180p } \frac{\alpha_1  T}{n}\right).\end{align*}
Therefore, by Markov inequality, we can conclude that there is no $S \subset V$ such that
$e(S,S) \ge \frac{p-q}{16}\frac{\alpha_1 T}{n}|S|$ and $|S|= \exp \left(-\frac{(p-q)^2 }{20p } \frac{\alpha_1
    T}{n}\right) n$ with high probability.

To conclude the proof of the lemma, we build the following sequence of sets. Let $\{ Z(i) \subset V\}_{1\le i \le i^{\star}}$ be generated as follows:
\begin{itemize}
\item $Z(0)=Z_1$.
\item For $i \ge 1$, $Z(i) = Z(i-1) \cup \{v_i \}$ if  there exists $v_i \in V$ such that $e(v_i , Z(i-1)) \ge \frac{p-q}{8}\frac{\alpha_1 T}{n}$ and $v_i \notin Z(i-1)$ and if there does not exist, the sequence ends.
 \end{itemize}
By construction, every $v \in V\setminus Z(i^\star)$ satisfies the conditions (H1) and (H2). Since $H$ is the largest set of which elements satisfy (H1) and (H2), $|H| \ge |V\setminus Z(i^\star)|$.   

The proof is hence completed if we show that $|Z(i^\star)|<  \exp
\left(-\frac{(p-q)^2 }{20p } \frac{\alpha_1 T}{n}\right) n$. Let
$t^{\star} =  \exp \left(-\frac{(p-q)^2 }{20p } \frac{\alpha_1
    T}{n}\right) n -|Z_1|$. If $i^{\star} \ge t^{\star},$
$|Z(t^{\star})| =  \exp \left(-\frac{(p-q)^2 }{20p } \frac{\alpha_1
    T}{n}\right) n$ and since $|Z_1 | \le \frac{1}{2} \exp \left(-\frac{(p-q)^2
  }{20p } \frac{\alpha_! T}{n}\right) n$,
$$e(Z(t^\star) , Z(t^\star))\ge \sum_{i=1}^{t^\star}e(v_i , Z(i-1)) \ge t^\star  \frac{p-q}{8}\frac{\alpha_1 T}{n} \ge \frac{|Z(t^\star )|}{2}\frac{p-q}{8}\frac{\alpha_1 T}{n},$$
However, from the previous claim, we know that with high probability,
all $S\subset V$ such that $|S| =  \exp \left(-\frac{(p-q)^2 }{20p }
  \frac{\alpha_1 T}{n}\right) n$ have to satisfy $e(S,S) <
\frac{p-q}{16}\frac{\alpha_1 T}{n}|S|$. Therefore, with high
probability, $i^\star < t^\star$ and
$$|Z(i^\star )|=i^{\star}+|Z_1| < t^{\star} + |Z_1| =  \exp \left(-\frac{(p-q)^2 }{20p } \frac{\alpha_1 T}{n}\right) n.$$

\subsection{Proof of Lemma~\ref{lem:improve}}

In this proof, we use the notation: $\mu(v, S) = \Ex [e(v,S) ].$ The numbers of
positive observations between a node $v\in H$ and the various clusters are concentrated
around their average due to condition (H1). So, at each improvement step,
nodes move to a cluster having more positive observations. We prove the lemma using the deviation of the number of positive observations of misclassified nodes (the deviation means the difference between the number of observations and its average). We derive a lower and an upper bound of the
deviation, and deduce an estimate of the fraction of misclassified nodes that become well-classified at each iteration.

Let $\mathcal{E}^{(i)} = \bigcup_{1\le  k \le K}(S^{(i)}_k
\setminus V_k)\cap H$. We first derive a lower bound for \\$\sum_{v\in \mathcal{E}^{(i+1)}} \sum_{k=1}^{K}\frac{|e(v,S^{(i)}_k)-\mu(v,S^{(i)}_k) |}{|
  S^{(i)}_k |}$. For all $v \in (S^{(i+1)}_a \cap V_b)\cap H$ and
$a\neq b$,
\begin{align*}
\sum_{k=1}^{K}\frac{|e(v,S^{(i)}_k)-\mu(v,S^{(i)}_k) |}{| S^{(i)}_k |}
&\ge \left|\frac{e(v,S^{(i)}_a)-\mu(v,S^{(i)}_a) }{|S^{(i)}_a |} - \frac{e(v,S^{(i)}_b)-\mu(v,S^{(i)}_b) }{| S^{(i)}_b |}\right| \cr
&\stackrel{(a)}{\ge}~\left|\frac{\mu(v,S^{(i)}_b) }{| S^{(i)}_b |} - \frac{\mu(v,S^{(i)}_a) }{| S^{(i)}_a |}\right|~\ge ~\frac{\alpha_1-\delta}{\alpha_1+\delta}(p-q)\frac{2T}{n^2},
\end{align*}
where $(a)$ stems from the facts that $v \in S^{(i+1)}_a$ satisfies $\frac{e(v,S^{(i)}_a) }{|
  S^{(i)}_a| } - \frac{e(v,S^{(i)}_b) }{| S^{(i)}_b | } >0$ and $v \in V_b$ satisfies
$\frac{\mu(v,S^{(i)}_b) }{| S^{(i)}_b |} - \frac{\mu(v,S^{(i)}_b) }{| S^{(i)}_b |} > 0$ (by design of the algorithm).
Therefore,
\begin{equation}
\sum_{v\in \mathcal{E}^{(i+1)}}\sum_{k=1}^{K}\frac{|e(v,S^{(i)}_k)-\mu(v,S^{(i)}_k) |}{|
  S^{(i)}_k |} \ge
\frac{\alpha-\delta}{\alpha+\delta}(p-q)\frac{2T}{n^2} |\mathcal{E}^{(i+1)}|. \label{eq:lbndimp}
\end{equation}

Now, we derive an upper bound of $\sum_{v\in \mathcal{E}^{(i+1)}}\sum_{k=1}^{K}\frac{|e(v,S^{(i)}_k)-\mu(v,S^{(i)}_k) |}{| S^{(i)}_k |}.$ To simplify the notation, we introduce\begin{eqnarray*}
Y_1(v)& = &\sum_{k=1}^{K}\frac{\left| e(v,V_k)-\mu(v,V_k) \right|}{|
  V_k |} , \cr
Y_2(v)& = &\sum_{k=1}^{K}\frac{\left| e(v, (S^{(i)}_k \setminus V_k) \cap
    H)-\mu(v,(S^{(i)}_k \setminus V_k) \cap H) \right|}{| V_k |} , \cr
Y_3(v)& = &\sum_{k=1}^{K}\frac{\left| e(v, (V_k \setminus S^{(i)}_k) \cap H)-\mu(v,(V_k \setminus S^{(i)}_k) \cap H) \right|}{| V_k |} , \cr
Y_4(v)& = &\sum_{k=1}^{K}\frac{\left| e(v,V\setminus H)-\mu(v,V\setminus H)
  \right|}{|V_k |} . 
\end{eqnarray*}
Then,
\begin{equation}
\sum_{v\in \mathcal{E}^{(i+1)}}\sum_{k=1}^{K}\frac{|e(v,S^{(i)}_k)-\mu(v,S^{(i)}_k) |}{| S^{(i)}_k |} \le \frac{\alpha_1}{\alpha_1-\delta}\sum_{v\in  \mathcal{E}^{(i+1)}} Y_1(v)+Y_2(v)+Y_3(v)+Y_4(v). \label{eq:ubndimp}
\end{equation}
When we combine \eqref{eq:lbndimp} and \eqref{eq:ubndimp}, we get
\begin{equation}
\frac{(\alpha_1-\delta)^2}{\alpha_1(\alpha_1+\delta)}(p-q)\frac{2T}{n^2} |
\mathcal{E}^{(i+1)} | \le \sum_{v\in  \mathcal{E}^{(i+1)}}
Y_1(v)+Y_2(v)+Y_3(v) + Y_4(v)
\label{eq:impbnd}
\end{equation}
From condition (H1), $v \in H$ satisfies
$\frac{|e(v,V_k)-\mu(v,V_k)|}{|V_k|} \le \frac{(p-q)2T}{n^2}.$ Thus,
$$\sum_{v\in  \mathcal{E}^{(i+1)}} Y_1(v) \le (p-q)\frac{T}{n^2} \cdot |  \mathcal{E}^{(i+1)}|.$$
From the definition of matrix $X = A - \mathbb{E}[A]$,
\begin{eqnarray*}
\sum_{v\in  \mathcal{E}^{(i+1)}} Y_2(v) & \le &\frac{\sum_{k=1}^K|
1_{ \mathcal{E}^{(i+1)}}^T \cdot X_{\Gamma} \cdot
1_{ (S^{(i)}_k\setminus V_k) \cap H}|}{\alpha_1 n}  \cr
& \le & \frac{\|1_{ \mathcal{E}^{(i+1)}}\|\cdot \|X_{\Gamma}\|\cdot \|1_{ \mathcal{E}^{(i)}}\|}{\alpha_1 n} \cr
&\le & \frac{ \|X_{\Gamma}\| \cdot \sqrt{ |  \mathcal{E}^{(i+1)}
    |}\sqrt{|  \mathcal{E}^{(i)}|}}{\alpha_1 n}.\end{eqnarray*}
Similarly, since $\bigcup_{k=1}^K(S^{(i)}_k\setminus V_k)\cap H =
\bigcup_{k=1}^K(V_k \setminus S^{(i)}_k)\cap H$,
\begin{eqnarray*}
\sum_{v\in  \mathcal{E}^{(i+1)}} Y_3(v) & \le &\frac{\sum_{k=1}^K|
1_{ \mathcal{E}^{(i+1)}}^T \cdot X_{\Gamma} \cdot
1_{ (V_k \setminus S^{(i)}_k) \cap H}|}{\alpha_1 n}  \cr
& \le & \frac{\|1_{ \mathcal{E}^{(i+1)}}\|\cdot \|X_{\Gamma}\|\cdot \|1_{ \mathcal{E}^{(i)}}\|}{\alpha_1 n} \cr
&\le & \frac{ \|X_{\Gamma}\| \cdot \sqrt{ |  \mathcal{E}^{(i+1)}
    |}\sqrt{|  \mathcal{E}^{(i)}|}}{\alpha_1 n}.\end{eqnarray*}
From condition (H2),
$$\sum_{v\in  \mathcal{E}^{(i+1)}} Y_4(v) \le
\frac{p-q}{4}\frac{T}{n^2}  |  \mathcal{E}^{(i+1)} |.$$
When we plug the above bounds for $Y_1 ,$ $Y_2 ,$ $Y_3 ,$ and $Y_4$ into
(\ref{eq:impbnd}), we conclude that 
\begin{eqnarray*}
\sqrt{\frac{ | \mathcal{E}^{(i+1)} |}{| \mathcal{E}^{(i)}|}}
& \le & \frac{\|X_{\Gamma} \|}{ \frac{\alpha_1 T}{n}(p-q)\left(\frac{(\alpha_1 - \delta)^2}{\alpha(\alpha_1 + \delta)} - \frac{5}{8}  \right) } 
\end{eqnarray*}

\section{Proof of Theorem~\ref{thm:adapt-results}}

We show that with high probability $\varepsilon (n,T) \le \exp\left(-\frac{T}{6n} \big( KL(p,q)
  + KL(q , p) \big) \right)$ under following conditions:
\begin{itemize}
\item[(C1)] $|S_k \setminus V_k| = 0$ for all $k.$
\item[(C2)] $(1-10^{-2}) (p-q) \le \hat{p} - \hat{q} \le (1+10^{-2})(p-q).$
\end{itemize}
The following lemma states that these conditinos hold with high probability.
\begin{lemma}
With high probability, (C1) and (C2) hold.\label{lem:condonadap}
\end{lemma}

In Algorithm~\ref{alg:adaptive}, we refer to as a classification process the procedure that attempts to classify a node in Step 4 (this procedure starts with "for $v\in R$"). For any $k$, $v \in V_k$ can be misclassified in Step 4 of the algorithm for two reasons: 1. $k^{\star}(v) \neq k$ and
$d^{\star}(v) \ge \frac{\hat{p}-\hat{q}}{2K}\frac{T}{n}$ (the node is assigned to a wrong cluster), and 2. $v$ is not assigned to any cluster. Refer to Algoirthm 4 for the precise definitions of $k^\star(v)$ and $d^\star(v)$.

We prove that with high probability, all nodes in $R$ are actually assigned to a cluster in one of the classification processes, and we provide an upper bound on the probability that $v$ is misclassified due to the first reason (reason 1.).

We first bound the probability that the classification processes end due to a lack of observation budget. Denote by $Y_v$ the number of required classification processes to assign $v$ to a cluster, assuming that there is no limit on the observation budget. After identifying the reference
kernels, at most $\frac{6}{5} n = (4T/5)/(2T/3n)$ classification processes can be run. Hence $\mathbb{P}\{ \sum_{v\in R} Y_v \le \frac{6}{5}n \}$ is the probability
that the classification processes end because all nodes are assigned to a cluster.

\begin{lemma}Under (C1) and (C2), for any classification process on $v \in {V}_k$,
$$\mathbb{P} \left\{ d^{\star}(v)  \le \frac{( \hat{p}-\hat{q})T}{2Kn}  \right\} \le 54K^2 \cdot (\frac{(p-q)^2}{p+q}\frac{T}{n})^{-1} .$$
\label{lem:ucheb}
\end{lemma}
From Lemma~\ref{lem:ucheb}, $Pr\{Y_v = i \} \le \eta^{i-1}
(1-\eta)$ for all $i \ge 2,$ where $\eta = 54K^2
(\frac{(p-q)^2}{p+q}\frac{T}{n})^{-1}.$ Hence, 
\begin{eqnarray*}
\Ex [ \exp (\sum_{v\in R}Y_v)] &=& \prod_{v\in R} \Ex [ \exp (Y_v)] \cr
 & \le & \left( \sum_{i \ge 1} (1-\eta) \eta^{i-1} e^i \right)^n\cr
 & \le &\left( \frac{(1-\eta) e}{1-\eta e} \right)^n ~=~\left(e+\frac{e(e-1)\eta}{1-   \eta e} \right)^n\cr
&\le&\exp\left(n+n\frac{(e-1)\eta }{1- \eta e} \right) ~ = ~ \exp\left( n\frac{1-\eta}{1- \eta e} \right).
\end{eqnarray*}
Applying Markov inequality,
\begin{align*}
\mathbb{P}\{\sum_{v\in R}Y_v \ge \frac{6}{5}n \} &\le \exp
\left(-\frac{6}{5}n +\frac{ 1-\eta}{1- \eta e}n \right) \cr 
& \le
\exp\left(-\frac{n}{10} \right),
\end{align*}
where the last inequality stems from the assumptions made in the theorem  ($\frac{(p-q)^2}{p+q}\frac{T}{n}= \Omega (1)$). Thus, with high probability, the classification of the remaining nodes is ended because there exists no more node.

The next lemma bounds the probability that a node is misclassified in a classification process.

\begin{lemma}Under (C1) and (C2), for any classification process on $v \in {V}_k$,
\begin{align*}\mathbb{P}& \left\{ k^{\star}(v) \neq k, d^{\star}(v)  \ge \frac{(\hat{p}-\hat{q})T}{2Kn}   \right\}  \le (K-1)\exp\left(-\frac{2T}{5Kn} \big( KL(p,q)
  + KL(q , p) \big) \right).\end{align*}\label{lem:upart}
\end{lemma}
Since the total number of the classification processes is at most $\frac{6}{5}n,$
$$\Ex [ \varepsilon(n,T)] \le  \frac{6}{5}(K-1)\exp\left(-\frac{2T}{5Kn} \big( KL(p,q)
  + KL(q, p) \big) \right).$$
Therefore, by Markov inequality, with high probability, 
$$ \varepsilon(n,T) \le  \exp\left(-\frac{T}{3Kn} \big( KL(p,q)
  + KL(q, p) \big) \right),$$
which concludes the proof of the theorem. It remains to establish the various intermediate lemmas.

\subsection{Proof of Lemma~\ref{lem:condonadap}}
To identify the reference kenels, we use $T/5$ budget on
$\frac{n}{5\log n}$ nodes. From the law of large number, with high probability, $\left| \frac{|S\cap V_k|}{|V_k|}
  -\frac{1}{5 \log n} \right| \le 10^{-2}$ for all $k$.
Therefore, from Theorem~\ref{thm:algorithms} and the assumptions made in the theorem: $\mathbb{E}[\varepsilon(n,T)] \le \frac{1}{n^2}$. Thus, by applying Markov inequality, we obtain that with high probability, 
$$\sum_{k}  | S_k \setminus V_k | =0.$$
Thus, with high probability, (C1) holds. Under condition (C1), by law of large number, it is easy to show
that (C2) holds with high probability.

\subsection{Proof of Lemma \ref{lem:ucheb}}

Since observations are independent, $S_k \subset V_k$ for all
$k$ (condition (C1) holds with high probability), and the number of
observations for each kernel is $\frac{2T}{3Kn}$ in a classification
process, for all $k' \neq k$,
\begin{eqnarray*}
\Ex [e(v,S_k)-e(v,S_{k'})]& = & \frac{2T}{3Kn}(p-q)\cr
\sigma[e(v,S_{k})-e(v,S_{k'})]^2& \le & \frac{2T}{3Kn}(p +q).
\end{eqnarray*}

By Chebyshev's inequality and condition (C2),
\begin{align*}
\mathbb{P}\left\{ d^{\star}(v)  \le  \frac{T (\hat{p}-\hat{q})}{2Kn} \right\}
&\le\sum_{k' \neq k} \mathbb{P} \left\{ e(v,S_k)-e(v,S_{k'})  \le  \frac{5T (p-q)}{9Kn} \right\} \cr
& \le\sum_{k' \neq k} \mathbb{P}\left\{ |e(v,S_k)-e(v,S_{k'})-\frac{2T}{3Kn}(p-q)| \ge \frac{T}{9Kn}(p-q)  \right\} \cr
& \le 54K^2 \cdot
(\frac{(p-q)^2}{p+q}\frac{T}{n})^{-1}.
\end{align*}

\subsection{Proof of Lemma \ref{lem:upart}}

Consider a classification process for $v\in V_k$. Let $\ell^+_v(t)$ be the outcome on the $t$-th observation for a pair $(v,w)$ with $w\in S_k$
and let $\ell^-_v(t)$ be the outcome of the $t$-th observation for a pair $(v,w)$ with $w\in S_{k'}$. Let $e(v,S_k)$ be the number of positive observations between node $v$ and cluster $S_k$, during the classification process. Then $e(v,S_k)- e(v,S_{k'} ) = \sum_{t=1}^{\frac{2T}{3Kn}}\ell^+_v
(t) -\sum_{t=1}^{\frac{2T}{3Kn}}\ell^-_v (t).$ Since $S_k \subset
V_k$ (from condition (C1)), $\mathbb{P}\{\ell^+_v (t) =1 \} =p$ and $\mathbb{P}\{\ell^-_v (t) =1 \} =q.$
Applying Markov inequality,
\begin{eqnarray}
\mathbb{P} \left\{ e(v,S_k) - e(v,S_{k'}) \le - \frac{( \hat{p} - \hat{q})T}{2Kn} \right\} &\le&
\frac{\Ex \left[ \exp \left( - \big( e(v,S_k) - e(v,S_{k'}) \big) \log \frac{p(1-q)}{q(1-p)} 
    \right)  \right]}{\exp\left( \frac{( \hat{p} - \hat{q})T}{2Kn} \log \frac{p(1-q)}{q(1-p)}\right)}
     \cr 
&= &\exp\left(-  \frac{(
    \hat{p} - \hat{q})T}{2Kn} \log \frac{p(1-q)}{q(1-p)} \right) \cr& \le& \exp\left(-  \frac{2(
    p - q)T}{5Kn} \log \frac{p(1-q)}{q(1-p)} \right), \label{eq:adapeebnd}
\end{eqnarray}
where the last inequality stems from condition (C2) and the second
equality comes from:
\begin{eqnarray*} 
\Ex \left[ \exp \left( e(v,S_1)\log \frac{q(1-p)}{p(1-q)} \right)  \right] &=&
\left(  \frac{1-p}{1-q}  \right)^{\frac{ 2T}{3Kn}}  \quad \mbox{and}  \cr
\Ex \left[ \exp \left( e(v,S_2)\log \frac{p(1-q)}{q(1-p)} \right)  \right] &=&
\left(  \frac{1-q}{1-p}  \right)^{\frac{ 2T}{3Kn}}, 
\end{eqnarray*}
because the $\ell^+_{v}(t)$'s are i.i.d for $1\le t \le \frac{T}{3n}$ and so are the $\ell^-_{v}(t)$'s for $1\le t \le \frac{T}{3n}$.

Finally, by combining Boole's inequality and \eqref{eq:adapeebnd},
\begin{align*}
\mathbb{P} \left\{ k^{\star}(v) \neq k, d^{\star}(v)  \ge
  \frac{(\hat{p}-\hat{q})T}{2Kn}   \right\}  &= \mathbb{P} \left\{
  e(v,S_k) - e(v,S_{k'}) \le - \frac{( \hat{p} -
    \hat{q})T}{2Kn}~\exists~k'\neq k \right\} \cr
&\le (K-1)\exp\left(-\frac{2T}{5Kn} \big( KL(p,q)
  + KL(q , p) \big) \right).
\end{align*}


\end{document}